\documentclass[11pt]{article}
\usepackage[english]{babel}
\usepackage[utf8]{inputenc}
\usepackage[T1]{fontenc}
\usepackage{amsmath}
\usepackage{amssymb}
\usepackage{graphicx}
\usepackage{subfigure}
\usepackage{psfrag}
\usepackage{color}

\allowhyphens

\newcommand{\fii}{\varphi}

\newcommand{\eps}{\varepsilon}

\newcommand{\ordo}{\mathcal{O}}

\newtheorem{thm}{Theorem}

\newcommand{\ket}[1]{{\left|#1\right\rangle}}
\newcommand{\bra}[1]{{\left\langle #1\right|}}

\newcommand{\skalarszorzat}[2]{{\langle #1 | #2 \rangle}}
\newcommand{\varhatoertek}[1]{\left\langle #1 \right\rangle}
\newcommand{\vev}{\varhatoertek}

\def\LY{Lee--\!Yang}
\newcommand{\One}{{\rm 1\!\!1}}
\newcommand{\ds}{\displaystyle}
\newcommand\B[3]{{\ensuremath{ {}^{\sscr(#1)}\!B_{#2}^{\sscr #3}}}}
\newcommand\C[3]{{\ensuremath{ C_{#1 #2}{}^{#3} }}}
\newcommand{\bea}{\begin{eqnarray}}
\newcommand{\eea}{\end{eqnarray}}
\newcommand{\be}{\begin{equation}}
\newcommand{\ee}{\end{equation}}
\newcommand{\sscr}{\scriptscriptstyle}
\newcommand{\fract}[2]{{\textstyle\frac{#1}{#2}}}
\newcommand{\m}{\phantom{-}}
\newcommand{\Ga}{\Gamma}
\newcommand{\hyp}{{\ensuremath{{}_2^{\vphantom\phi}{\rm F}_{\!1}^{\vphantom\phi}}}}
\newcommand{\D}{{{\rm d}}}

\renewcommand{\vec}[1]{|\,#1\,\rangle}
\newcommand{\cev}[1]{\langle \,#1\,|}
\newcommand{\veev}[2]{\langle\,#1\,|\,#2\,\rangle}
\def\hc{\hat h_c}
\def\hhc{{{\hat h}_{\rm crit}}}

\usepackage{ifpdf}

\ifpdf
\usepackage{epstopdf}
\usepackage[pdftex]{hyperref}
\else
\usepackage[hypertex,ps2pdf,dvips,colorlinks,urlcolor=blue,citecolor=blue,linkcolor=blue]{hyperref}
\fi

\usepackage{babel}
\makeatother

\begin{document}

\thispagestyle{empty}

~ \vspace{0.5cm}

\begin{center} \vskip 14mm
{\Large\bf One-point functions in massive integrable QFT with boundaries}\\[20mm] 
{\large 
M\'arton Kormos$^{a,b}$ and Bal\'azs Pozsgay$^{c}$~~\footnote{Emails: 
{\tt kormos@sissa.it, pozsgay.balazs@gmail.com}}}
\\[8mm]
\it$^a$ International School for Advanced Studies (SISSA),\\
Via Beirut 4, 34014 Trieste, Italy
\\[8mm]
$^b$ Istituto Nazionale di Fisica Nucleare, Sezione di Trieste, Italy
\\[8mm]
$^c$ Institute for Theoretical Physics, Universiteit van Amsterdam,\\
 Valckenierstraat 65, 1018 XE Amsterdam, The Netherlands
\vskip 22mm
\end{center}

\begin{quote}{\bf Abstract}\\[1mm]
We consider the expectation value of a local operator on a strip with
non-trivial boundaries in 1+1 dimensional massive integrable
QFT. Using finite volume regularisation in the crossed channel and
extending  the boundary state formalism to the finite volume case
we give a series expansion
for the one-point function in terms of the exact form factors of the
theory. The truncated series is compared with the numerical results of
the truncated conformal space approach in the scaling \LY\ model. We
discuss the relevance of our results to quantum quench problems.

\end{quote}

\vfill
\newpage

\numberwithin{equation}{section}

\tableofcontents

\section{Introduction}

Finite size effects play a central role in quantum field theory and
statistical physics. Apart from having direct relevance to statistical
physics models in finite volume, describing for example boundary
critical phenomena or percolation problems, they naturally appear in
the description of systems at finite temperature. For example,
two-dimensional Euclidean field theories with a finite, periodic
direction provide a framework for studying one-dimensional theories at
finite temperature. Moreover, boundary phenomena and finite volume
systems can play an important role in the understanding of quantum
quenches: in certain cases the boundaries play the role of the initial
and final states of the non-equilibrium problem
\cite{Calabrese:2006rx,demler-quench,davide}. In addition, in many
cases, especially for numerical simulations, the system under
consideration is put in a finite volume box. In this case it is
essential to understand the finite size behaviour of various
quantities.

Correlation functions are very important both at finite and infinite
volume because they encode a lot of non-trivial information about the
spectrum and the interactions in the theory. Calculating correlators is
not a simple task even in integrable theories. In 1+1 dimensional
integrable quantum field theories the form factor approach provides an
efficient way to calculate correlation functions at zero temperature
and in infinite volume. The essence of the method is inserting a
complete set of asymptotic states into the correlators and then making
use of the explicit forms of the appearing matrix elements of the
operators, the so-called form factors, which are known in many
integrable models.

However, for finite temperature or equivalently, for finite geometry
the applicability of the form factor approach at present is somewhat
constrained. There has been considerable progress in free theories
(for example the quantum Ising model), where the non-interacting
nature of the theory allows for calculating correlation functions both in
the finite temperature setting \cite{sachdev} or in the presence of a
boundary \cite{Konik:1995ws,esslerIsing}. However, interacting
theories pose technical and conceptual difficulties, which have not
yet been overcome.

In the finite temperature case one particular approach was developed
by LeClair and Mussardo \cite{leclair-mussardo}. They proposed an
integral series for correlation functions based on the exact form
factors and the Thermodynamic Bethe Ansatz equations.  Their result
for one-point functions were checked in particular examples
\cite{fring,saleurfiniteT,mussardo2}, and then a highly non-trivial
check was given by confirming it up to the third order in the
low-temperature expansion using finite volume regularisation
\cite{fftcsa2}. For two-point functions, however, some counterexamples
were found where the formalism does not seem to work
\cite{fring,saleurfiniteT} and the problem is still far from being
settled.

In this work we address the generalised problem of one-point
functions in finite volume where instead of periodic boundary
conditions we consider non-trivial boundary conditions, thus we have a
strip geometry instead of a cylinder. In this sense our work can be a
first step in finding an expression similar to the LeClair--Mussardo
formula in the boundary case. Expectation values in the presence of
boundaries in general depend on the position of the operator in
question\footnote{This has to be contrasted with the problem of
  expectation values of boundary operators considered for example in
  \cite{Takacs:2008ec}.}. The exact determination of them is thus
non-trivial; for one boundary the problem has comparable difficulty to
that of the two-point functions in infinite volume; and similarly, in
the presence of two boundaries the technical problems resemble the
case of a three-point function or a two-point function at finite
temperature.

In the work \cite{gerard} the authors considered the case of one
boundary, or in other words the problem when the operator is much
closer to one of the boundaries than the volume of the
system. 
They developed a form factor expansion which was later used in
\cite{samaj-2005-72,demler-quench} to study certain problems in
condensed-matter systems. In addition, the analytic results were
confronted in \cite{gerard} with the numerical Truncated
  Conformal Space Approach (TCSA). 
Our expansion reproduces and goes
beyond their result, and we will show that at small enough volume our
results give a remarkable improvement. Our numerical comparison with
the TCSA results is very similar to their method. On the other hand,
the theoretical determination of the spectral series is more involved:
there appear conceptual (and also technical) difficulties, which are
not present in \cite{gerard}. Most importantly, one has to deal with
certain singularities of the form factors and also the divergent
contributions to the partition function. In this work we make use of a
finite volume regularisation scheme, which first was used in
interacting field theory in \cite{fftcsa1,fftcsa2} and later in
\cite{Essler:2007jp,essler-2009}.

As remarked above, integrable boundary field theory can be used to
investigate quench problems. Recent papers \cite{demler-quench,davide}
consider the time evolution of one-point functions after certain types
of global quenches. As we will show, their problem and method have
similarities to ours and thus our results also have relevance to
quench problems. 

The outline of the paper is the following. In section 2 we introduce
the finite volume boundary states and as a first application we give a
series expansion of the finite volume ground state energy. In section
3 we turn to the problem of one-point functions. After describing the
method in detail we present the explicit calculations of the first
terms in the form factor expansion. At the end of the section we
discuss the relation of our expression with existing results in the
literature and we also explain its connection to quench
problems. Section 4 is devoted to the numerical comparison between the
spectral expansion and the TCSA data for the scaling \LY\ model,
considered as a perturbed conformal field theory. First we collect the
relevant properties and formulae of the model, then we discuss briefly
the TCSA method and how one-point functions can be obtained in this
framework. Finally we compare the form factor and TCSA results for
various combinations of boundary conditions and strip widths. Our
conclusions are given in section 5. For some technical details and for
a collection of our final result the reader is referred to the
appendices.

\section{Boundary states and expectation values}

For the sake of simplicity let us consider an integrable relativistic
field theory with only one particle species with mass $m$ and
two-particle S-matrix $S(\theta)$. We are interested in the case where
the theory lives on a finite line segment of length $R$ with
boundary conditions at the edges that do not spoil the integrability.

If $B$ denotes such an integrable boundary then the scattering of an
incoming multi-particle state\footnote{The subscript $B$ indicates
  that the state is in the Hilbert space of the boundary theory.},
\be
|A(\theta_1)A(\theta_2)\dots A(\theta_n)\rangle_B\;,
\ee
is purely elastic, i.e. the set of outgoing rapidities is
$\{-\theta_1,-\theta_2,\dots,-\theta_N\}$ and the scattering can be described by
a product of one-particle reflection amplitudes which are defined as \cite{Ghoshal:1993tm}
\be
|A(\theta)\rangle_B=R_B(\theta)|A(-\theta)\rangle_B\;.
\ee

\begin{figure}
  \centering
\psfrag{hab}{$\hspace{0.2cm}H^{\alpha,\beta}_R$}
\psfrag{HL}{$H_L$}
\psfrag{alpha}{$\alpha$}
\psfrag{beta}{$\beta$}
\psfrag{Balpha}{$\ket{B_\alpha^L}$}
\psfrag{Bbeta}{$\ket{B_\beta^L}$}
\psfrag{R}{$R$}
\psfrag{L}{$L$}
  \subfigure[
]{\includegraphics[scale=0.4]{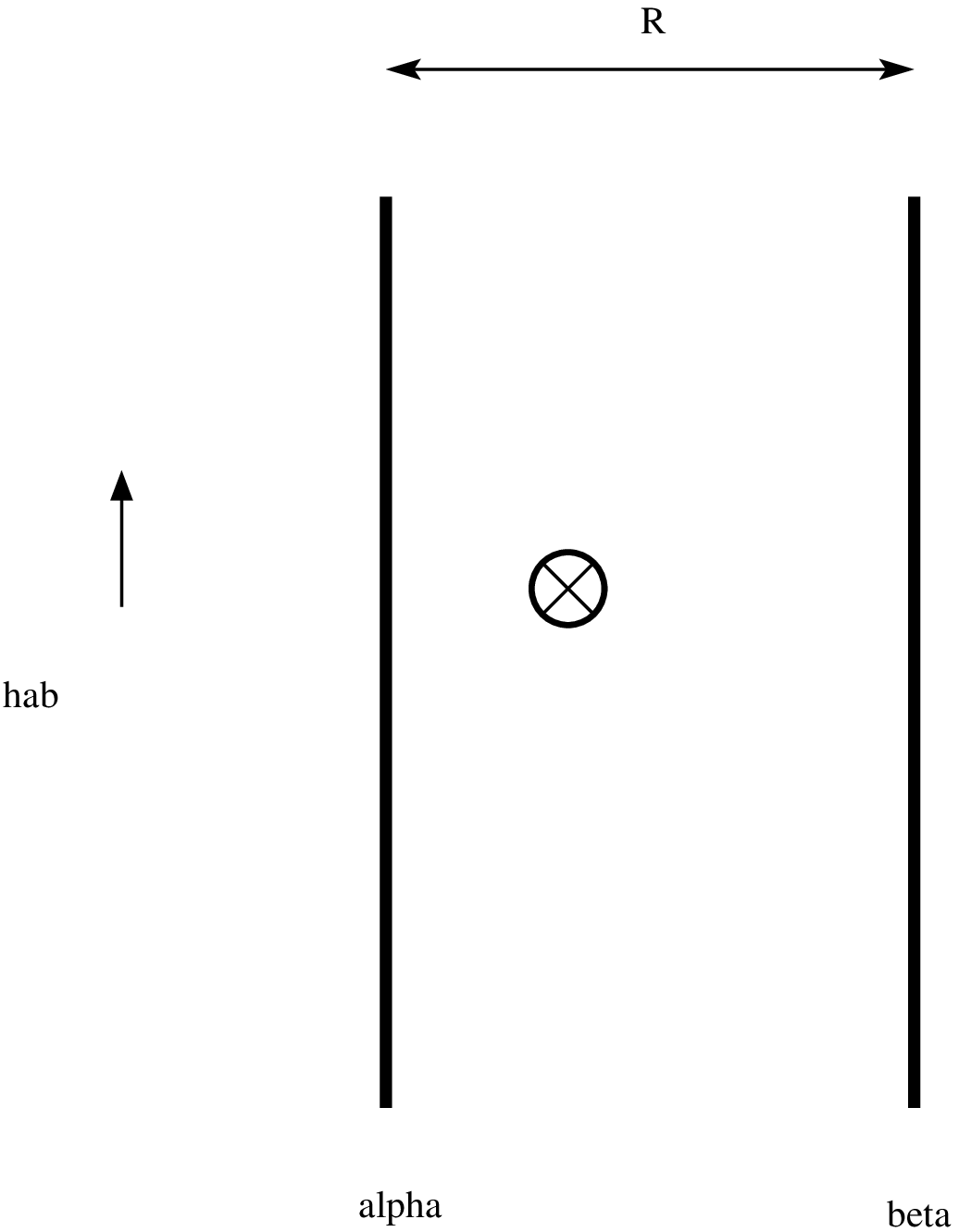}\label{alap1}}
  \subfigure[
]{\includegraphics[scale=0.4]{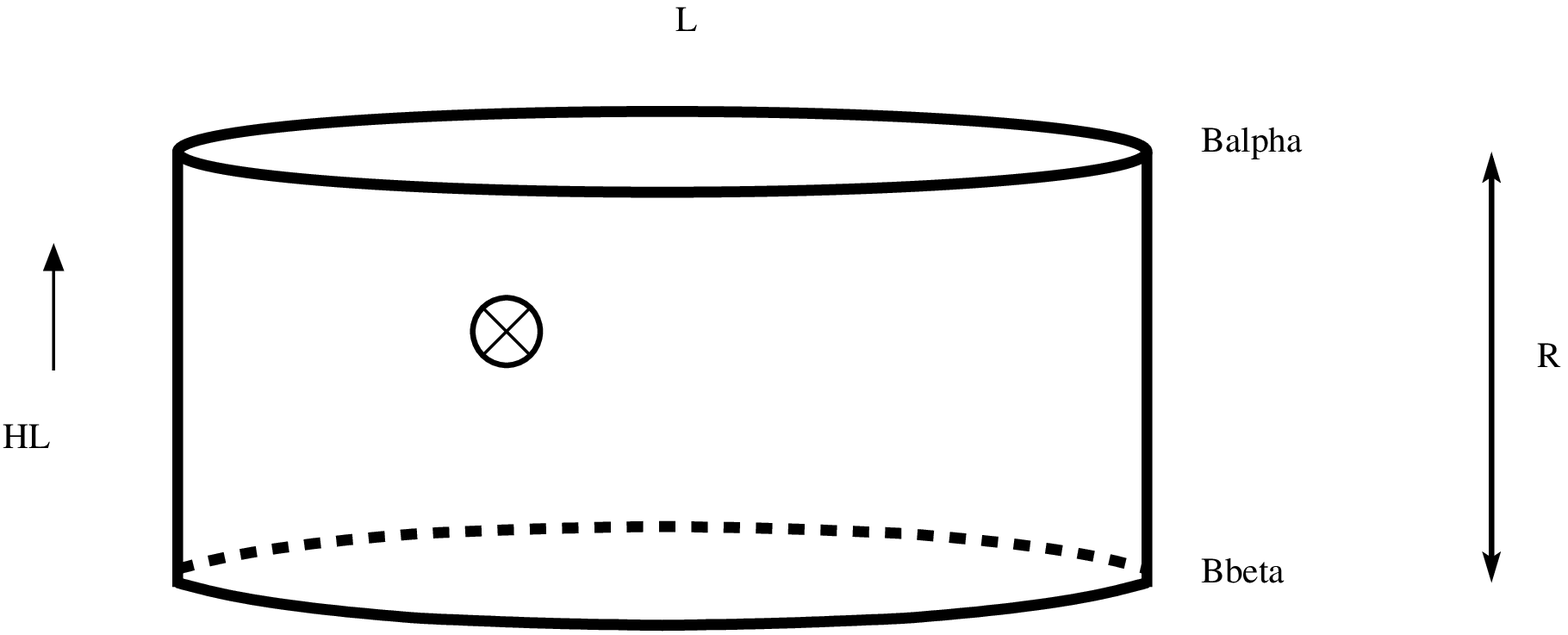}\label{alapL}}
\caption{One-point function in the original strip geometry (a) and
    the finite volume
  regularisation performed in the crossed
  channel (b). The upward arrows denote the (imaginary) time evolution
  generated by the corresponding Hamiltonians. }
\end{figure}

The main objective of this paper is to evaluate the vacuum expectation value 
\begin{equation}
\label{objective}
  \vev{O(x)}_R^{\alpha,\beta} 
\end{equation}
in the presence of the integrable boundaries $\alpha$ and
$\beta$ where $x\in [0,R]$. Here the expectation value is taken with
respect to the vacuum state of the finite volume Hamiltonian
$H^{\alpha,\beta}_R$ (see fig. \ref{alap1}). In a more general
setting one can also take excited states corresponding to boundary
bound states but we do not elaborate on this case here.

One can consider the same
quantity \eqref{objective} after a Euclidean rotation. In this picture $R$ plays the
role of the Euclidean time variable and the expectation value is given
by the formal expression
\begin{equation}
\label{ezkellene}
\vev{O(x)}_R^{\alpha,\beta}=
\frac{\bra{B_\alpha}\ e^{-H x}O\ 
e^{-H(R-x)}\ket{B_\beta}}{\bra{B_\alpha}e^{-HR}\ket{B_\beta}}\;,
\end{equation}
where $H$ is the infinite volume Hamiltonian
and $\ket{B_\alpha}$ and $\ket{B_\beta}$ are the boundary states
corresponding to the boundary conditions $\alpha$ and $\beta$, respectively. 
When they do not contain zero-momentum particles, they can be expanded
in the asymptotic multi-particle basis
as \cite{Ghoshal:1993tm}
\begin{equation}
\label{boundarystate1}
  \ket{B_j}=\mathcal{N}_j \exp\left(\int \frac{d\theta}{4\pi}
      K_j(\theta) A(-\theta)A(\theta) \right) \ket{0}\;,
\end{equation}
where
\begin{equation*}
  K_j(\theta)=R_j(i\pi/2-\theta)\;,\qquad j=\alpha,\beta
\end{equation*}
and $A(\theta)$ are the Faddeev--Zamolodchikov creation operators
satisfying the commutation relations
\begin{equation*}
  A(\theta_1)A(\theta_2)=S(\theta_1-\theta_2)A(\theta_2)A(\theta_1)\;.
\end{equation*}
The integrals run from $-\infty$ to $\infty$ unless otherwise stated.
The amplitudes $K_i(\theta)$ satisfy the ``boundary cross-unitary
equation'' \cite{Ghoshal:1993tm}
\begin{equation*}
  K_j(\theta)=S(2\theta)K_j(-\theta)\;,
\end{equation*}
which serves as a consistency relation of \eqref{boundarystate1}. The
behaviour under complex conjugation is given by $K_j(\theta)^*=K_j(-\theta)$.

In general there may be additional contributions to the boundary state
involving zero-momentum particles, which can be associated with a
non-zero coupling of a single particle state to the boundary in
  the original channel. In this case there is a corresponding pole in
the reflection factor at $\theta=i\pi/2$:
\begin{equation*}
  R_j(\theta)\sim \frac{i}{2} \frac{(g_j)^2}{\theta-i\pi/2}\;,
\end{equation*}
and the boundary state is given by
\begin{equation}
\label{Bstate}
  \ket{B_j}=\mathcal{N}_j \exp\left(\bar{g}_jA(0)+\int \frac{d\theta}{4\pi}
 K_j(\theta) A(-\theta)A(\theta)\right) \ket{0}\;,
\end{equation}
where $\bar{g}_j$ represents the one-particle coupling to the boundary.
Originally  it was argued that $\bar{g_j}=g_j$ \cite{Ghoshal:1993tm}, however
it was found numerically in \cite{gerard} that the proper
normalisation is  $\bar{g_j}=g_j/2$. This claim was later proven on
 general
grounds in \cite{Bajnok:2004tq,Bajnok:2006dn}.  
The normalisation constants $\mathcal{N}_j$ are infinite in the
  infinite volume system. However, this is not a serious problem as they
  drop out from the calculation of the vacuum expectation values and
  one may set them to unity. 

The general strategy to evaluate \eqref{ezkellene} now is the
following. We substitute expression \eqref{Bstate} for the boundary
states and expand the exponentials. The resulting multi-particle
states are eigenstates of the Hamiltonian, and the matrix elements of
the local operator $O$ between these states are the form factors,
which can be determined in principle and which are actually known in
many integrable models.

The special case for the $R\to\infty$ limit of \eqref{ezkellene} with $x$ fixed (one-point
function in the presence of only one boundary) was investigated in
\cite{gerard}. In this case there are no conceptual and technical
  difficulties and the spectral expansion described above (valid for $g_\alpha=0$) reads
\begin{equation}
  \label{eq:only_x}
  \vev{{O}(x)}
=\sum_{n=0}^\infty \frac{1}{2^n n!} \int \frac{d\theta_1}{2\pi}\dots \frac{d\theta_n}{2\pi}
F_{2n}^{{O}}(-\theta_1,\theta_1,\dots,-\theta_n,\theta_n)
\prod_{i=1}^n \Big(K_\alpha(\theta_i) e^{-2m\cosh\theta_i x} \Big)\;,
\end{equation}
where
\begin{equation*}
  F_m^{{O}}(\theta_1,\dots,\theta_m)=\bra{0}\mathcal{O}\ket{\theta_1,\dots,\theta_m}
\end{equation*}
are the elementary form factors of $O$ in infinite volume.
In integrable models they are meromorphic functions of the rapidities
which satisfy the so-called form factor axioms (Watson equations),
which can be
considered as axioms for the form factor bootstrap. Supplied with
the principles of maximum analyticity and the cluster property they
contain enough information to determine the form factors
completely. In a theory with only one particle species the form factor
axioms read:

I. Lorentz invariance:
\begin{equation}
   F_{n}^{{O}}(\theta_{1}+\Lambda,\dots,\theta_{n}+\Lambda)=
e^{s\Lambda} F_{n}^{{O}}(\theta_{1},\dots,\theta_{n})\;,
\end{equation}
where $s$ is the Lorentz spin of the operator $O$. In this work we
only consider scalar operators corresponding to $s=0$.

II.
 Exchange:
\begin{equation}
  F_{n}^{{O}}(\theta_{1},\dots,\theta_{k},\theta_{k+1},\dots,\theta_{n})=
  S(\theta_{k}-\theta_{k+1})
F_{n}^{{O}}(\theta_{1},\dots,\theta_{k+1},\theta_{k},\dots,\theta_{n})\;.
\label{eq:exchangeaxiom}
\end{equation}

III.
 Cyclic permutation: 
\begin{equation}
F_{n}^{O}(\theta_{1}+2i\pi,\theta_{2},\dots,\theta_{n})=F_{n}^{O}(\theta_{2},\dots,\theta_{n},\theta_{1})\label{eq:cyclicaxiom}\;.
\end{equation}

IV.
 Kinematical singularity
\begin{equation}
-i\mathop{\textrm{Res}}_{\theta=\theta^{'}}
F_{n+2}^{O}(\theta+i\pi,\theta^{'},\theta_{1},\dots,\theta_{n})=
\left(1-\prod_{k=1}^{n}S(\theta-\theta_{k})\right)F_{n}^{O}(\theta_{1},\dots,\theta_{n})\;.
\label{eq:kinematicalaxiom}\end{equation}

V. Dynamical singularity 
\begin{equation}
-i\mathop{\textrm{Res}}_{\theta=\theta^{'}}F_{n+2}^{\mathcal{O}}(\theta+iu,\theta^{'}-iu,\theta_{1},\dots,\theta_{n})=\Gamma
F_{n+1}^{\mathcal{O}}(\theta,\theta_{1},\dots,\theta_{n})\;,\label{eq:dynamicalaxiom}
\end{equation}
where $\Gamma$ is the on-shell three-particle coupling corresponding to a bound
state pole of the S-matrix
\begin{equation}
S(\theta\sim
i2u)\sim\frac{i\Gamma^{2}}{\theta-i2u}\;,\quad\theta\sim 2iu\;.\label{eq:Smatpole}
\end{equation}
In a theory with only one particle, like the scaling \LY\ model the only possibility is $u=\pi/3$.

We also note that all form factors can be expressed in terms of the elementary form factors
with the help of the crossing relation 
\begin{multline*}
F_{mn}^{O}(\theta_{1}^{'},\dots,\theta_{m}^{'}|\theta_{1},\dots,\theta_{n})=
 F_{m-1,n+1}^{{O}}(\theta_{1}^{'},\dots,\theta_{m-1}^{'}|\theta_{m}^{'}+i\pi,\theta_{1},\dots,\theta_{n})
 \\
 +\sum_{k=1}^{n}\Big(2\pi\delta(\theta_{m}^{'}-\theta_{k})\prod_{l=1}^{k-1}S(\theta_{l}-\theta_{k})\times
 F_{m-1,n-1}^{{O}}(\theta_{1}^{'},\dots,\theta_{m-1}^{'}|\theta_{1},\dots,\theta_{k-1},\theta_{k+1}\dots,\theta_{n})\Big)\;.
\end{multline*}
For an introduction to the form factor bootstrap program we refer the
reader to the review of Smirnov \cite{smirnov_ff} and to papers
\cite{karowski_weisz_elso_S,zam_Lee_Yang,delfino_mussardo,ising_ff1}.

In the simultaneous presence of the two boundaries the expression
\eqref{ezkellene} is ill-defined.
There are divergent contributions both to the numerator and the
denominator, which are associated with disconnected
terms of the form factors and with various contributions to the
partition function. 
One way to evaluate \eqref{ezkellene} is to introduce finite volume as
a regulator of the singular contributions. This method was successfully
used in \cite{fftcsa2} to derive a rigorous low-temperature expansion for
one-point functions at finite temperature, which in the crossed
channel corresponds to vacuum
expectation values in a finite volume with periodic boundary
conditions. In this sense the present work is an extension of
the approach of \cite{fftcsa2} to the more general setting of arbitrary (integrable) boundary conditions.

We introduce a large finite volume $L$ in the crossed channel (with periodic boundary
conditions) and consider the limit
\begin{equation}
\label{ezkell}
\vev{O(x)}_R=\lim_{L\to\infty}   \vev{O(x)}_R^L =
\lim_{L\to\infty}
\frac{\Big\langle {B_\alpha^L}\Big|\ e^{-H_L x}O(0,0)\ 
e^{-H_L(R-x)}\ket{B_\beta^L}}{\Big\langle {B_\alpha^L}\Big| e^{-H_LR}\ket{B_\beta^L}}\;,
\end{equation}
where  $H_L$ is the finite volume Hamiltonian and 
$\ket{B_j^L}$ represent the boundary states in finite volume (see
fig. \ref{alapL}). It will
be the subject of the next subsection to properly expand them in the basis of
the eigenstates of $H_L$. 

\subsection{Boundary states in finite volume}

The boundary states can be expanded in finite volume as
\begin{equation*}
  \ket{B_j^L}=\sum_\Psi G_j^\Psi(L) \ket{\Psi}_L\;,\qquad j=\alpha,\beta\;,
\end{equation*}
where $\ket{\Psi}_L$ are eigenstates of the finite volume Hamiltonian
$H_L$. In particular the function $G_j^0(L)$ determines the large $R$
behaviour of the partition function and can be written as
\begin{equation*}
  G_j^0(L)=e^{-f_jL}g^0_j(L)\;.
\end{equation*}
Here $f_j$ is the contribution of a single boundary to the ground state energy $E_0(R)$
and $g^0_j(L)$ is the standard non-perturbative
$g$-function\footnote{This should not
  be confused with the one-particle boundary coupling $g_j$.} which was
introduced in critical systems by Affleck and Ludwig
\cite{Affleck:1991tk}. In theories with only massive excitations in
the bulk the $g$-function decays
exponentially. The exact integral series for
the $g$-function in massive theories was derived in \cite{Dorey:2004xk}; for a
first treatment of a non-trivial massless flow see \cite{Dorey:2009vg}.

The excited states $\ket{\Psi}_L$ of a finite volume
Hamiltonian can be described in a large volume
as scattering states consisting of $n$ particles with rapidities
$\theta_n$ given by the solution of the asymptotic Bethe--Yang
equations
\begin{equation}
Q_k=mL\sinh\theta_k +\sum_{j\ne k} \delta(\theta_k-\theta_j)=2\pi
  I_k\;,\quad\quad k=1,\dots,n\;.
\label{eq:BY}
\end{equation}
It is convenient to define
a continuous two-particle phase shift function by\footnote{
  We made use of the fact that the effective statistics is fermionic,
  i.e.\ $S(0)=-1$. All known integrable models possess this property, the
only counter-example being the free boson. In a theory with more than
one particle species the most convenient choice for the definition of
the phase shift is $S_{ab}(\theta)=S_{ab}(0)e^{i\delta_{ab}(\theta)}$.}
\begin{equation*}
  S(\theta)=-e^{i\delta(\theta)}\;,\qquad\qquad \delta(-\theta)=-\delta(\theta)\;.
\end{equation*}
This poses the following prescription for the momentum quantum numbers:
\begin{equation*}
\begin{split}
  I_k \in \mathbb{Z} \qquad \text{for odd }n\;,\qquad\qquad
I_k \in \mathbb{Z}+\frac{1}{2}  \qquad \text{for even }n\;.
\end{split}
\end{equation*}
The quantum numbers $\{I_1,\dots,I_n\}$ completely determine the
individual scattering states, therefore they may be used to label the
sates as
\begin{equation*}
  \ket{\{I_1\dots I_n\}}_L \;, 
\end{equation*}
where the states are normalised to unity:
\begin{equation*}
_L\skalarszorzat{\{I_1\dots I_n\}}{\{I'_1\dots I'_m\}}_L=\delta_{nm}
  \delta_{I_1,I'_1}\dots \delta_{I_n,I'_n}\;.
\end{equation*}
The total energy and momentum are calculated additively as
\begin{equation*}
  E=\sum_{i=1}^n m\cosh\theta_i+\ordo(e^{-\mu L})\;,\qquad\qquad
 P=\sum_{i=1}^n m\sinh\theta_i+\ordo(e^{-\mu L})\;.
\end{equation*}
The exponential corrections are governed by the mass scale $\mu$ which
is uniform in the sense that it is determined by the analytic
structure of the S-matrix and it does not depend on the particular
multi-particle state in question. For a systematic treatment of
exponential corrections to excitation energies see
\cite{luscher_1particle,klassen_melzer} and also recent papers
\cite{Bajnok:2008bm,Hatsuda:2008na}. In a massive field theory it is
expected that the leading exponential corrections of more complicated
quantities (like form factors \cite{Pozsgay:2008bf} and correlation
functions) are also of order $e^{-\mu L}$.

For later use we introduce the density of states (in 
rapidity-space) in the $n$-particle
sector of the theory, which is given by the Jacobian
\begin{equation}
  \rho_n(\theta_1,\dots,\theta_n)=\det J_{kl}\;,\qquad
  J_{kl}=\frac{\partial Q_k}{\partial\theta_l}\;.
\end{equation}

Apart from normalisation issues the boundary states in finite and infinite
volume should have the same structure. Therefore, it is natural to 
expand the finite volume boundary state up to the
two-particle contribution as
\begin{equation}
\label{finiteL_kifejtes}
\begin{split}
& \ket{B_j^L}=G_j^0(L)\Big(\ket{0}_L+\frac{g_i}{2}N_1(L)\ket{\{0\}}_L
+\sum_{I>0} K_j(\theta) N_2(\theta,L) \ket{\{-I,I\}}_L+\dots \Big)\;,
\end{split}
\end{equation}
where in the last term it is understood that the rapidities $\theta$
are the solutions of the appropriate Bethe--Yang equation
\begin{equation}
\label{barQ1}
  \bar{Q}_1(\theta)=mL\sinh\theta+\delta(2\theta)=2\pi I\;.
\end{equation}
 Note also
that the sum in \eqref{finiteL_kifejtes} only runs over $I>0$ because the states
$\ket{\{I,-I\}}_L$ and $\ket{\{-I,I\}}_L$ are identical. In
\eqref{finiteL_kifejtes} we introduced the normalisation factors $N_1(L)$
and $ N_2(\theta,L)$ which are not determined by first principles. 
In the following we calculate them including  all finite size effects which scale as negative powers of
$L$ (only neglecting those which decay exponentially).
To this order we consider
vacuum expectation values in the $R\to\infty$ limit with 
fixed $x$, 
both in a large finite volume $L$ and directly in the infinite system.

In infinite volume the first three
contributions read
\cite{gerard}
\begin{equation}
\label{infty_kifejtes_1}
\vev{{O}(x)}_\alpha=\vev{{O}}+\frac{g_\alpha}{2} F_1^O e^{-mx}+
\frac{1}{2}\int \frac{d\theta}{2\pi} K_\alpha^*(\theta) 
F_2^O(\theta,-\theta)e^{-2mx\cosh\theta}+\dots\;.
\end{equation}
The same quantity can also be evaluated in finite volume
as
\begin{equation}
\label{finiteL_kicsit_konkretabb}
  \vev{{O}(x)}_\alpha^L=
\vev{{O}} + 
\frac{g_\alpha}{2}N_1(L) \bra{\{0\}}{O}\ket{0}_L e^{-E_0 x}+
\sum_{I>0} K^*_\alpha(\theta)N_2(\theta,L)  \bra{\{-I,I\}}{O}\ket{0}_L
e^{-E_I x}+\dots\;,
\end{equation}
where 
\begin{equation*}
  E_0=m+\mathcal{O}(e^{-\mu L})\;,\qquad\qquad
E_I=2m\cosh\theta+\mathcal{O}(e^{-\mu L})\;,
\end{equation*}
and the finite volume form factors are given by \cite{fftcsa1}
\begin{equation*}
   \bra{\{0\}}{O}\ket{0}_L=\frac{F_1^O}{\sqrt{mL}}+\mathcal{O}(e^{-\mu L})\;,\qquad\quad 
 \bra{\{-I,I\}}{O}\ket{0}_L
 =\frac{F_2^O(\theta,-\theta)}{\sqrt{\rho_2(\theta,-\theta)}}
+\mathcal{O}(e^{-\mu L})\;.
\end{equation*}
The summation in \eqref{finiteL_kicsit_konkretabb} can be replaced by
an integration leading to
\begin{multline}
\label{finiteL_kicsit_konkretabb2}
  \vev{{O}(x)}_\alpha^L
=\vev{{O}} + 
\frac{g_\alpha}{2} \frac{N_1(L)}{\sqrt{mL}} F_1^O e^{-m x}+\\
+\frac{1}{2}\int \frac{d \theta}{2\pi} 
\frac{\bar\rho_1(\theta)N_2(\theta,L)}{\sqrt{\rho_2(\theta,-\theta)}}
K^*_\alpha(\theta)  F_2^O(\theta,-\theta)e^{-2mx\cosh\theta }
+\mathcal{O}(e^{-3m x})+\mathcal{O}(e^{-\mu  L})\;,
\end{multline}
where $\bar\rho_1(\theta)$ is the constrained density of states given by 
\begin{equation*}
  \bar\rho_1(\theta)=\frac{d\bar Q_1}{d\theta}=mL\cosh\theta+2\varphi(2\theta)\;.
\end{equation*}
In a massive theory
\begin{equation*}
    \vev{{O}(x)}_\alpha-  \vev{{O}(x)}_\alpha^L\quad \sim\quad
    \mathcal{O}(e^{-\mu L})\;.
\end{equation*}
Requiring that expressions \eqref{infty_kifejtes_1} and
\eqref{finiteL_kicsit_konkretabb2} only differ by exponentially small
terms for any scalar operator yields
\begin{equation}
\label{N1N2}
  N_1(L)=\sqrt{mL}+ \mathcal{O}(e^{-\mu L})\;, 
\qquad\qquad N_2(\theta,L)=\frac{\sqrt{\rho_2(\theta,-\theta)}}{\bar\rho_1(\theta)}+ \mathcal{O}(e^{-\mu L})\;.
\end{equation}
Note that this argument essentially coincides with that used in
\cite{fftcsa1} to determine the normalisation factors of the
elementary finite volume form factors.

The two-particle density satisfies
\begin{equation*}
  \rho_2(\theta,-\theta)=\rho_1(\theta)\bar\rho_1(\theta)\;,
\end{equation*}
therefore
\begin{equation}
\label{N2maskepp}
  N_2(\theta,L)=\sqrt{\frac{\rho_1(\theta)}{\bar\rho_1(\theta)}}
=1-\frac{\varphi(2\theta)}{mL\cosh\theta}+\mathcal{O}(1/L^2)\;. 
\end{equation}
We stress that the eqs. \eqref{N1N2} include all finite
volume corrections which behave as negative powers of $L$, in
particular there can be no additional $\mathcal{O}(1/L)$ corrections to $N_2$. This
will become important in the evaluation of vacuum expectation values
at finite $R$.

It is straightforward to extend the argument above to higher
excited states. The finite volume boundary state is given up to the
four-particle terms by
\begin{multline}
\label{valami}
 \ket{B_j^L}=G_j^0(L)\Big(\ket{0}_L+\frac{g_j}{2}\sqrt{mL}\ket{\{0\}}_L
+\sum_I K_j(\theta) N_2(\theta,L) \ket{\{I,-I\}}_L \\
+\sum_I \frac{g_j}{2} K_j(\theta) N_3(\theta,L) \ket{\{I,-I,0\}}_L
+\frac{1}{2}\mathop{\sum_{IJ}}_{I\ne J}  
K_j(\theta_1) K_j(\theta_2) N_4(\theta_1,\theta_2,L) \ket{\{I,-I,J,-J\}}_L\Big)\;.
\end{multline}
The three-particle sector consists of states with rapidities $\{-\theta,0,\theta\}$
where $\theta$ is determined by the single quantisation condition
\begin{equation}
\label{barQ3}
  \bar Q_3(\theta)=mL\sinh\theta+\delta(\theta)+\delta(2\theta)=2\pi I\;.
\end{equation}
Therefore the normalisation of the three-particle states is given by 
\begin{equation*}
   N_3(\theta)=\frac{\sqrt{\rho_3(\theta,0,-\theta)}}{\bar\rho_3(\theta)}\;,
\end{equation*}
where
\begin{equation*}
  \bar\rho_3(\theta)=\frac{d\bar Q_3(\theta)}{d\theta}=mL\cosh\theta+\varphi(\theta)+2\varphi(2\theta)\;.
\end{equation*}

In the four-particle case the quantisation condition for the rapidities
$\{-\theta_1,\theta_1,-\theta_2,\theta_2\}$ is given by the
system of equations
\begin{equation*}
  \begin{split}
    \bar Q_{4,1}&=mL\sinh\theta_1+\delta(\theta_1-\theta_2)+\delta(\theta_1+\theta_2)+\delta(2\theta_1)=
2\pi I_1\;,\\
   \bar Q_{4,2}&=mL\sinh\theta_2+\delta(\theta_2-\theta_1)+\delta(\theta_1+\theta_2)+\delta(2\theta_2)=
2\pi I_2\;,\\
  \end{split}
\end{equation*}
yielding the normalisation 
\begin{equation*}
  N_4(\theta_1,\theta_2)=\frac{\sqrt{\rho_4(\theta_1,-\theta_1,\theta_2,-\theta_2)}}
{\bar\rho_4(\theta_1,\theta_2)}\;,
\end{equation*}
where
\begin{equation*}
  \bar\rho_4(\theta_1,\theta_2)=\det J \qquad\text{with}\qquad J_{ik}=\frac{\partial
    \bar Q_{4,i}}{\partial \theta_k}\;,\quad\quad i,k=1,2\;.
\end{equation*}

\subsection{Large volume expansion of the Casimir energy}

\label{long-distance}

Before turning to the evaluation of the one-point function we consider
the Casimir energy in the two-boundary setting. 
In
\cite{LeClair:1995uf} the exact non-perturbative value was found to be
\begin{equation}
  \label{eq:BTBA_E0}
  E_0(R)=-\frac{1}{2}\int \frac{d\theta}{2\pi}
  m\cosh\theta \log\Big(1+K_\alpha^*(\theta)K_\beta(\theta)e^{-\eps(\theta)}\Big)\;,
\end{equation}
where $\eps(\theta)$ is the solution of the boundary TBA equation
\begin{equation}
  \label{eq:BTBA}
  \eps(\theta)=2mR\cosh\theta-\int \frac{d\theta}{2\pi}
  \varphi(\theta-\theta')\log\Big(1+K_\alpha^*(\theta')K_\beta(\theta')e^{-\eps(\theta')}\Big)\;,
\end{equation}
where $\varphi(\theta)=\mathrm{d} \delta(\theta)/\mathrm{d}
  \theta$. Expression \eqref{eq:BTBA_E0} is
  normalised to have the asymptotics $E_0(R)\to 0$ as $R\to
  \infty$. The actual ground state energy also includes bulk and
  boundary contributions, which can be calculated from the boundary
  TBA \cite{Us1}.

Equations \eqref{eq:BTBA_E0}-\eqref{eq:BTBA} were derived  in
\cite{LeClair:1995uf} for the case $g_\alpha=g_\beta=0$ 
 but it was argued that they yield the correct result even in the presence
of zero-momentum particles. Although this claim was supported 
by several numerical checks \cite{Us1}, a rigorous proof is still
missing. Moreover, it turned out that there are certain cases when
\eqref{eq:BTBA_E0}-\eqref{eq:BTBA} cannot be 
true.

In \cite{Bajnok:2004tq,Bajnok:2006dn,Bajnok:2007ep} it was shown that
if $g_\alpha\ne 0 \ne g_\beta$ then 
the leading contribution in the integral in \eqref{eq:BTBA_E0} is given by
\begin{equation}
  \label{eq:E0_rossz}
  E_0(R)=-m\frac{\big|g_\alpha g_\beta\big|}{4}e^{-mR}+\dots\;.
\end{equation}
On the other hand, the authors argued that
the leading behaviour of the Casimir energy is always
\begin{equation}
  \label{eq:E0_leading}
  E_0(R)=-m\frac{g_\alpha g_\beta}{4}e^{-mR}+\dots\;,
\end{equation}
irrespective of the sign of $g_\alpha g_\beta$. This discrepancy shows that
the validity of \eqref{eq:BTBA_E0} is restricted to the case
$g_\alpha g_\beta>0$. On the other hand, it was shown in
\cite{Bajnok:2004tq,Bajnok:2006dn,Bajnok:2007ep} that it is possible to derive
a proper analytic continuation  to the region
$g_\alpha g_\beta<0$ with the correct leading behaviour
\eqref{eq:E0_leading}. 

Here we derive a large $R$ expansion of the Casimir energy by means of
the finite volume regularisation of the partition function. Our
results will be compared to \eqref{eq:BTBA_E0}.  We follow closely the
treatment of \cite{Bajnok:2004tq,Bajnok:2006dn}, however we are able
to extend their method to obtain the second order terms of the Casimir
energy. In order to keep the exposition simple, we only consider the
region $g_\alpha g_\beta>0$ where the BTBA
\eqref{eq:BTBA_E0}-\eqref{eq:BTBA} is applicable.

The partition function $Z(L,R)$ can be evaluated in two different channels.
Treating $L$ as time variable and $R$ as space one obtains in the
large $L$ limit
\begin{equation*}
  Z=e^{-E_0(R)L}\left(1+\mathcal{O}(e^{-mL})\right)\;,
\end{equation*}
where $E_0(R)$ is the ground-state energy in a finite volume $R$. It
follows that
\begin{equation*}
  E_0(R)=-\lim_{L\to\infty} \frac{\log Z}{L}\;.
\end{equation*}
Treating $R$ as time variable (corresponding to thermal field theory
with temperature $T=1/R$) the partition function can be developed into
a low-temperature expansion. In the regime $mL e^{-mR}\ll 1$ it is
sufficient to sum over the low-lying excitations:
\begin{equation}
\label{Zeloszor}
  Z=1+N_1^2 \frac{g_\alpha g_\beta}{4} e^{-mR}+\sum_I \big(N_2(\theta)
  \big)^2 K_\alpha^*(\theta)K_\beta(\theta) e^{-2mR\cosh\theta}+\mathcal{O}(e^{-3mR})\;.
\end{equation}

Note that in the expression above we did not perform a complex
conjugation for $g_\alpha$, although the amplitude $K_\alpha(\theta)$
is conjugated. The motivation for this prescription is a delicate
issue. In unitary theories the one-particle couplings are always real,
therefore it is only the non-unitary case which has to be addressed.
In non-unitary theories $g_\alpha$ is purely imaginary, therefore
complex conjugation simply results in changing the sign of the
appropriate term. On the other hand, the space of the states is not
positive-definite, in fact one-particle states can be assigned a
negative norm. This additional change of sign accounts for the correct
prescription in \eqref{Zeloszor}. A different argument is provided by
the corresponding conventions in CFT, where the prescription for the
inner product does not include complex conjugation. Moreover, when
comparing our form factor expansion with TCSA results in section
\ref{sec:compare} we will show that this choice of the sign is the
correct one.

If $mL\gg 1$ one can safely substitute the normalisations \eqref{N1N2}
into \eqref{Zeloszor}. The logarithm of the partition function is then
written as
\begin{multline}
\label{ZN1N2}
\frac{\log Z}{L}
=m\frac{g_\alpha g_\beta}{4} e^{-mR} +
\sum_I
  \frac{m\cosh\theta}{\bar\rho_1(\theta)} 
K_\alpha^*(\theta)K_\beta(\theta) e^{-2mR\cosh\theta}\\
-\frac{1}{2}m^2L\left(\frac{g_\alpha g_\beta}{4}
  e^{-mR}\right)^2+\ordo(e^{-\mu L})+\ordo(e^{-3mR})\;.
\end{multline}
The first term is already in complete accordance with
\eqref{eq:E0_leading}; in the following we also evaluate 
the second order terms. 
First of 
all note that the summation over $I$ is divergent because of the double pole
\begin{equation*}
 K_\alpha^*(\theta)K_\beta(\theta)\approx
 \frac{g_\alpha^2g_\beta^2}{4} \frac{1}{\theta^2}\;,\qquad
\theta\approx 0\;.
\end{equation*}
In order to determine the leading singularity
it suffices to use the lowest order approximation in \eqref{eq:BY}
\begin{equation*}
  \theta_I=\frac{2\pi I}{mL}\;, \qquad\qquad I=\frac{1}{2},\frac{3}{2},\frac{5}{2},\dots\;.
\end{equation*}
The divergent part of the summation in \eqref{ZN1N2} can then be
expressed as
\begin{equation*}
e^{-2mR}  \sum_I \frac{g_\alpha^2g_\beta^2}{4L} \left(\frac{mL}{2\pi
      I}\right)^2=m^2L\, e^{-2mR} \frac{g_\alpha^2g_\beta^2}{32}\;,
\end{equation*}
where we used the identity
\begin{equation*}
  \left(\frac{1}{2}\right)^2+ \left(\frac{3}{2}\right)^2+
  \left(\frac{5}{2}\right)^2+\dots=
\frac{\pi^2}{2}\;.
\end{equation*}
Therefore the $O(L e^{-2mR})$ terms cancel exactly in \eqref{ZN1N2}, as needed
to obtain a meaningful $L\to\infty$ limit. Note that this result
provides new evidence for the normalisation condition
$\bar{g}_j=g_j/2$, because the $L\to\infty$ limit would not exist with
a different choice of $\bar{g}_j$.

In the following we also evaluate the finite left-over piece of the
$\mathcal{O}(e^{-2mR})$ terms in \eqref{ZN1N2}. In order to make
calculations as transparent as possible we first introduce the
following theorem:

\begin{thm}
Let $\theta_I$ be the solutions of the quantisation condition
\begin{equation*}
   \bar{Q}_1(\theta)=mL\sinh\theta+\delta(2\theta)=2\pi I
\end{equation*}
and let $f(\theta)$ be a
   symmetric function which apart from a double pole at $\theta=0$ is
   analytic in a neighbourhood of the real axis:
\begin{equation*}
  f(\theta)\approx \frac{G}{\theta^2} \quad \text{as}\quad
    \theta\to 0\;.
\end{equation*}
Then the expression
\begin{equation*}
  S(L)=\left(\sum_I \frac{f(\theta_I)}{\bar\rho_1(\theta_I)}\right)-\frac{G}{8}mL
\end{equation*}
has a regular behaviour at large $L$ with the $L\to\infty$ limit given by
\begin{equation*}
  \lim_{L\to\infty} S(L)=I_f+K_f\;,
\end{equation*}
where
\begin{equation*}
\begin{split}
  I_f= \int_{-\infty}^\infty \frac{d\theta}{4\pi} \left(
f(\theta)-G\frac{\cosh\theta}{\sinh^2\theta}
\right)\qquad\text{and}\qquad
K_f=\frac{G}{4} \varphi(0)\;.
\end{split}
\end{equation*}
\end{thm}
The proof can be extracted from the Appendix B of
\cite{Takacs:2009fu}. In addition, we also present a 
different derivation in Appendix \ref{Prooff}.
In the present case Theorem 1 can be applied with the substitutions
\begin{equation*}
  f(\theta)=
  \cosh\theta K_\alpha(-\theta)K_\beta(\theta) e^{-2mR\cosh\theta}\;,
\qquad\qquad
G=\frac{g_\alpha^2g_\beta^2}{4} e^{-2mR}\;.
\end{equation*}
Here we made use of the identity
$K_\alpha^*(\theta)=K_\alpha(-\theta)$. It is easy to see that with
this choice $f(\theta)$ is indeed symmetric and it is analytic apart
from the prescribed double pole at $\theta=0$.
\label{thmsec}

The results of Theorem 1 for the Casimir energy yields
\begin{multline}
\label{E0R-e}
    E_0(R)=
-m\frac{g_\alpha g_\beta}{4} e^{-mR}-m\frac{g_\alpha^2g_\beta^2}{4} e^{-2mR} \frac{\varphi(0)}{4}\\
-\frac{1}{2}    \int \frac{d\theta}{2\pi}
  m\cosh\theta
  \left(K_\alpha(-\theta)K_\beta(\theta)e^{-2mR\cosh\theta} -
  \frac{g_\alpha^2g_\beta^2}{4\sinh^2\theta}e^{-2mR} 
\right)  +\ordo(e^{-3mR})\;.
\end{multline}

In order to compare this result to the BTBA equations it is
necessary to derive a large $R$ expansion of \eqref{eq:BTBA_E0}. First
of all we regularise the integral along the lines of  \cite{Bajnok:2004tq}:
\begin{equation}
 \label{BTBA_koztes1}
  E(R)=-m\frac{g_\alpha
    g_\beta}{4}\sqrt{e^{-\eps(0)}}-\frac{1}{2}\int\frac{d\theta}{2\pi}
m\cosh\theta
\log\left(\frac{1+K_\alpha(-\theta)K_\beta(\theta)e^{-\eps(\theta)}}
{1+\frac{g_\alpha^2g_\beta^2}{4\sinh^2\theta}e^{-\eps(0)}}\right)\;.
\end{equation}
Using the same trick for the integral in \eqref{eq:BTBA} it is possible 
to derive the leading correction to the pseudo-energy: 
\begin{equation}
  \label{eq:BTBA_tricky}
   \eps(\theta)=2mR\cosh\theta-\frac{g_\alpha g_\beta}{2}
   \varphi(\theta)e^{-mR\cosh\theta}+\ordo(e^{-2mR})\;,
\end{equation}
and in particular
\begin{equation}
  \sqrt{e^{-\eps(0)}}=e^{-mR}\Big(1+\frac{g_\alpha g_\beta}{4}
   \varphi(0)e^{-mR}+\dots\Big)\;.
\end{equation}
The second term in \eqref{BTBA_koztes1} can be expanded as
\begin{equation}
\begin{split}
&\log\left(1+\frac{K_\alpha(-\theta)K_\beta(\theta)e^{-\eps(\theta)}-
\frac{g_\alpha^2g_\beta^2}{4\sinh^2\theta}e^{-\eps(0)}} 
{1+\frac{g_\alpha^2g_\beta^2}{4\sinh^2\theta}e^{-\eps(0)}}
\right)=\\
&= \frac{K_\alpha(-\theta)K_\beta(\theta)e^{-\eps(\theta)}-
\frac{g_\alpha^2g_\beta^2}{4\sinh^2\theta}e^{-\eps(0)}} 
{1+\frac{g_\alpha^2g_\beta^2}{4\sinh^2\theta}e^{-\eps(0)}}
\approx \left(K_\alpha(-\theta)K_\beta(\theta)e^{-\eps(\theta)}-
\frac{g_\alpha^2g_\beta^2}{4\sinh^2\theta}e^{-\eps(0)} \right)\;.
\end{split}
\end{equation}
In the last step we used the fact that the extra term in the
denominator only makes a difference for $\theta \sim e^{-mR}$,
therefore it only contributes to the higher order terms.
In the last expression we may finally substitute the zeroth order
approximation $\eps(\theta)\approx 2mR\cosh\theta$.
Putting everything together we arrive at
\begin{multline}
  \label{eq:BTBA_full_2order}
  E_0(R)=
-m\frac{g_\alpha g_\beta}{4}e^{-mR}
-m\frac{g_\alpha^2 g_\beta^2}{16}\varphi(0)e^{-2mR}\\
-\frac{1}{2}\int\frac{d\theta}{2\pi}m\cosh\theta
\left(K_\alpha(-\theta)K_\beta(\theta)e^{-2mR\cosh(\theta)}-
\frac{g_\alpha^2g_\beta^2}{4\sinh^2\theta}e^{-2mR} \right)+\ordo(e^{-3mR})\;.
\end{multline}
This is in complete accordance with \eqref{E0R-e} and can be regarded as
a strong confirmation of the consistency of our calculations. Apart
from checking the identity $\bar{g}_j=g_j/2$ and also the validity
of the BTBA equations, we also confirmed the normalisation factor $N_2(\theta,L)$. 
In particular, a different $\ordo(1/L)$ term in $N_2(\theta,L)$ would
yield a different pre-factor for the term $\varphi(0)e^{-2mR}$.

Finally we note that our finite volume regularisation scheme is not
sensitive to the sign of $g_\alpha g_\beta$, in particular one arrives
at the same result \eqref{eq:BTBA_full_2order} also for $g_\alpha
g_\beta<0$. On the other hand, in this regime one has to use the
modified BTBA equations given by eq.\ (2.7) of \cite{Bajnok:2006dn}. We
leave it as an exercise to show that a careful large $R$ expansion of
this modified BTBA indeed reproduces \eqref{eq:BTBA_full_2order}.

\subsection{Connection between boundary states and  form factors}

As a remark to this section we would like to point out an
interesting similarity between the elementary 
finite volume form factors $\bra{0}O\ket{\Psi}_L$
and the 
amplitudes $g_j^{\Psi}(L)$ defined by
\begin{equation*}
  \skalarszorzat{B_j^L}{\Psi}_L=e^{-f_jL}g_j^\Psi(L)\;,\qquad j=\alpha,\beta\;,
\end{equation*}
where $f_j$ is the contribution of a single boundary to the ground
state energy on a strip. Both quantities measure the overlap of a
normalised eigenstate $\ket{\Psi}_L$ with a non-normalisable extended
state, namely the boundary state $\ket{B_j^L}$ and the state
$O\ket{0}_L$ created by acting with a local operator on the
vacuum. To demonstrate the analogy we compare the main results in the
case of the specific two-particle state
$\ket{\Psi}_L=\ket{\{I,-I\}}_L$. We have seen that
\begin{equation*}
  g_j^{\Psi}(L)=
K_j(\theta)  N_2(\theta,L)g_j^{0}(L) +\ordo(e^{-\mu L})\;.
\end{equation*}
The vacuum amplitude $g_j^{0}(L)$ only include exponentially small
corrections (see below), therefore the excited state amplitude can be
written as
\begin{equation*}
  g_j^{\Psi}(L)=
K_j(\theta)  N_2(\theta,L) +\ordo(e^{-\mu L})\;. 
\end{equation*}
On the other hand, the finite volume form factor was shown in
\cite{fftcsa1} to be given by
\begin{equation*}
\bra{0}O\ket{\Psi}_L=\frac{F_2^O(\theta,-\theta)}{\sqrt{\rho_2(\theta,-\theta)}}
+\ordo(e^{-\mu L})\;.
\end{equation*}
We observe that apart from the exponential corrections both objects are determined
by the corresponding infinite volume quantity and a normalisation
factor which only depends on the finite volume density of states. 
In the following we make some comments about the possible structure of
the exponential corrections. 

First of all it is instructive to recall the known results
in the case of $\ket{\Psi}_L=\ket{0}_L$, which corresponds to the
exact non-perturbative $g$-function and to the finite volume vacuum
expectation value  $\vev{O}_L$, respectively. 
In the case of the non-perturbative $g$-function
the final result was expressed in \cite{Dorey:2004xk} as 
\begin{multline}
\label{exact-g}
  2\log g_j^0(L)
=\frac{1}{2} \int \frac{d\theta}{2\pi} 
\Big(\Phi_j(\theta)-\delta(\theta)-2\varphi(2\theta) \Big)
\log\Big(1+e^{-\eps(\theta)}\Big)\\
+\sum_{i=1}^\infty \frac{1}{n}
\int \frac{d\theta_1}{2\pi}\dots \int \frac{d\theta_n}{2\pi}
\left(\prod_{i=1}^n \frac{1}{1+e^{\eps(\theta_i)}}\right)
\varphi(\theta_1+\theta_2)\varphi(\theta_2-\theta_3)\dots \varphi(\theta_n-\theta_1)\;,
\end{multline}
where $\Phi_j(\theta)=-i \frac{d}{d\theta}\log(R_j(\theta))$ and
$\eps(\theta)$ is the solution of the \textit{periodic-boundary-conditions} TBA equation 
\begin{equation*}
 \eps(\theta)=mL\cosh\theta-\int \frac{d\theta}{2\pi}
  \varphi(\theta-\theta')\log\Big(1+e^{-\eps(\theta')}\Big)\;.
\end{equation*}
In the case of the vacuum expectation value the relevant exact result
is the LeClair--Mussardo series \cite{leclair-mussardo} 
\begin{equation*}
  \varhatoertek{\mathcal{O}}_L^{\text{periodic}}=
\sum_{n=0}^\infty \frac{1}{n!} 
\int \frac{d\theta_1}{2\pi}\dots \int \frac{d\theta_n}{2\pi}
\left(\prod_{i=1}^n \frac{1}{1+e^{\eps(\theta_i)}}\right)
F_{2n,c}^\mathcal{O}(\theta_1,\dots,\theta_n)\;,
\end{equation*}
where $\eps(\theta)$ is the same pseudo-energy function as above. 
There is a striking structural similarity between the two series, most
importantly the weight functions appearing in the 
integrals are exactly the same. 

Based on these similarities
we conjecture that the structure of exponential corrections to the
excited state quantities  $\bra{0}O\ket{\Psi}_L$ and  $g_j^\Psi(L)$
will be similar to each other as well.
It was remarked in \cite{Dorey:2009vg} that the exact
series for the amplitudes $g_j^\Psi(L)$ probably involves the solution of
the corresponding excited state TBA. The appearance of the excited
state TBA was also observed in preliminary studies of exponential
corrections to form factors \cite{vajonlesz-evalami}. The exact result
for  $g_j^\Psi(L)$ will probably involve an integral series similar to
\eqref{exact-g} normalised by a ``dressed'' form of
$N_2(\theta,L)$.

\section{Evaluation of the vacuum expectation value}

In this section we develop a large $R$ expansion of the vacuum expectation
value (v.e.v.) $\vev{O}_R^{\alpha,\beta}$. We use the finite volume
regularisation scheme described in the previous section and express
the expectation value as
\begin{equation}
\label{tervszerint}
\vev{O}_R^L=\frac{\Big\langle B_\alpha^L\Big|\ e^{-H_L x}O\ 
e^{-H_L(R-x)}\ket{B_\beta^L}}{\Big\langle {B_\alpha^L}\Big| e^{-H_LR}\ket{B_\beta^L}}\;.
\end{equation}
We will calculate explicitly the v.e.v. with the boundary
states truncated to contributions with up to four
particles. In order to simplify notations we define
\begin{equation*}
  \ket{B_j^L}=\sum_{n=0}^4 \ket{B_j^L}^{(n)}\;,\qquad j=\alpha,\beta\;,
\end{equation*}
where
\begin{equation}
\begin{split}
\ket{B_j^L}^{(0)}&=\ket{0}_L\;,\\
\ket{B_j^L}^{(1)}&=\frac{g_j}{2}N_1(L)\ket{\{0\}}_L\;,\\
\ket{B_j^L}^{(2)}&=\sum_I K_j(\theta) N_2(\theta,L) \ket{\{I,-I\}}_L\;,\\
\ket{B_j^L}^{(3)}&=\sum_I \frac{g_j}{2} K_j(\theta) N_3(\theta,L) \ket{\{I,-I,0\}}_L\;,\\
\ket{B_j^L}^{(4)}&=\frac{1}{2}\mathop{\sum_{IJ}}_{I\ne J}  
  K_j(\theta_1) K_j(\theta_2) N_4(\theta_1,\theta_2,L) \ket{\{I,-I,J,-J\}}_L\;.
\end{split}
\end{equation}
We also define 
\begin{equation*}
  C_{nm}\,=\,^{(n)}\bra{B_\alpha^L}e^{-H_Lx}O(x)e^{-H_L(R-x)}\ket{B_\beta^L}^{(m)}\;.
\end{equation*}

In the following we perform a double expansion of
\eqref{tervszerint} with expansion parameters  $e^{-mx}$ and
$e^{-m(R-x)}$.
 In order to keep track 
of the various terms
let us introduce two auxiliary variables $u$ and $v$.  They are use to
count orders
of $e^{-mx}$ and $e^{-m(R-x)}$
and at the end of the calculations both are set to 1. Then the v.e.v.
takes the form 
\begin{equation}
\label{complete}
  \vev{O}_R^L=\frac{1}{Z}\sum u^n v^m C_{nm}= \sum u^n v^m \tilde{D}_{nm}\;.
\end{equation}
We define
\begin{equation*}
  Z=\sum_n (uv)^n Z_n\;,
\end{equation*}
where the first terms are given by
\begin{equation}
 Z_0=1\;,\qquad\qquad Z_1=\frac{g_\alpha g_\beta}{4} mL e^{-mR}\;,
\end{equation}
and
\begin{equation}
  Z_2=\sum_I  e^{-2m\cosh\theta R}
  K_\alpha^*(\theta)K_\beta(\theta) N_2(\theta,L)^2\;.
\end{equation}
The inverse of the partition function is expanded as
\begin{equation*}
  Z^{-1}=\sum_n (uv)^n \bar{Z}_n\;,
\end{equation*}
where the first few terms read
\begin{equation*}
  \bar{Z}_0=1\;,\qquad \bar{Z}_1=-Z_1\;,\qquad \bar{Z}_2=Z_1^2-Z_2\;.
\end{equation*}
Putting this together we obtain
\begin{equation}
\label{dnm}
  \tilde{D}_{nm}=\sum_l  C_{n-l,m-l}\bar{Z}_l\;,
\end{equation}
where the first few non-trivial terms are given by
\begin{align*}
  \tilde{D}_{1m}&=C_{1m}-Z_1C_{0,m-1}\;,  &m=1,2,\dots\;,\\
\tilde D_{2m}&=C_{2m}-Z_1 C_{1,m-1}+(Z_1^2-Z_2)C_{0,m-2}\;, &m=2,3,\dots\;.
\end{align*}

We expect that the quantities $\tilde D_{nm}$ have a regular
behaviour as $L\to\infty$ and for the actual limit we define
\begin{equation}
\label{dnm-uj}
    D_{nm}=\lim_{L\to\infty} \tilde D_{nm}\;.
\end{equation}
The vacuum expectation value is then expressed as
\begin{equation*}
  \vev{O}_R^{\alpha,\beta}=\sum_{n,m} D_{nm}\;.
\end{equation*}
Note that the individual terms contributing to $\tilde D_{nm}$ may contain
divergent pieces which scale with positive powers of $L$. The most
singular terms in the $n$ particle sector carry a factor of $(mL
e^{-mR})^n$, therefore the
expansion is valid in the regime 
\begin{equation*}
  1\ll mL \ll e^{mR}\;.
\end{equation*}
The $L\to\infty$ limit of the complete series  \eqref{complete} is to
be understood as an analytic
continuation. 

The evaluation of the individual $C_{nm}$ is built on the knowledge of
the finite volume form factors. In the general case they are given by
\cite{fftcsa1}
\begin{equation}
\label{fftcsalap}
\begin{split}
  \bra{\{I_1,\dots,I_n\}}O\ket{\{J_1,\dots,J_m\}}_L=
\frac{F^O_{n+m}(\theta_1+i\pi,\dots,\theta_n+i\pi,\theta'_1,\dots,\theta'_m)}
{\sqrt{\rho_n(\theta_1,\dots,\theta_n)\rho_m(\theta'_1,\dots,\theta'_m)}}
+\ordo(e^{-\mu L})\;,
\end{split}
\end{equation}
where it is understood that the rapidities
$\{\theta_1,\dots,\theta_n\}$ and  $\{\theta'_1,\dots,\theta'_m\}$
are solutions to the corresponding Bethe--Yang equations. Formula
\eqref{fftcsalap} is valid whenever there are no coinciding rapidities
in the two scattering states. In \cite{fftcsa1} it was shown that
the only two situations when coinciding rapidities occur are the case of
diagonal form factors and matrix elements between parity-symmetric
states containing zero-momentum particles. The general rule for
evaluating disconnected contributions can be found in
\cite{fftcsa2}; in the present work we will cite the necessary results
whenever they are needed.

Before turning to the actual calculations we have to address a very
important question concerning the phase of the form factors. Note that
equation \eqref{fftcsalap} is to be understood up to a phase factor;
in particular the order of the rapidities is not determined by first
principles. In fact, an additional phase factor corresponds to a
redefinition of the basis vectors, and this phase drops out from the
calculation of some physical quantities, for example correlation
functions in infinite volume. However, phase factors are utterly
relevant in the present case because they do affect the final result
for the v.e.v. A guideline can be established from the known results
in infinite volume, where the correct phase is fixed by consistency
arguments. First of all, the two-particle contributions to the
boundary state are written as
\begin{equation*}
K_j(\theta)A(-\theta)A(\theta)\;.
\end{equation*}
This expression is symmetric in $\theta$ as required by
consistency. 
When it comes to the
evaluation of the v.e.v. in the presence of a single
boundary, the above definition yields the expressions
\begin{equation*}
\dots  K_\beta(\theta) F_n^O(-\theta,\theta,\dots) \quad \text{ and } \quad
\dots K_\alpha^*(\theta) F_n^O(\theta+i\pi,-\theta+i\pi,\dots)\;,
\end{equation*}
where the dots stand for possible additional rapidities and amplitudes.
Observe that both expressions above are symmetric in $\theta$, therefore we apply the
following rule: whenever there appear the amplitudes
$K_\beta(\theta_i)$ and $K_\alpha^*(\theta_i)$ with some $\theta_i$,
the explicit form and order of the rapidities substituted into the relevant form factor
is given by $(-\theta_i,\theta_i)$ and  $(\theta_i+i\pi,-\theta_i+i\pi)$,
respectively. The exchange of two pairs of rapidities
does not make a difference, therefore the phase of the form factors is
completely fixed by the above rule. 

The complex conjugation of the amplitude $K_\alpha$ can be avoided by
means of the identity
\begin{equation*}
  K_\alpha^*(\theta_i)=K_\alpha(-\theta_i)\;.
\end{equation*}
An additional change of variables $\theta_i\to -\theta_i$ then amounts to
the convention
\begin{equation*}
\dots K_\alpha(\theta) F_n^O(-\theta+i\pi,+\theta+i\pi,\dots)\;.
\end{equation*}
Note also that the presence
of zero-momentum particles does not produce any additional
ambiguities, because it is only the expressions
\begin{equation*}
  K_j(\theta)F_n^O(-\theta,\theta,0,\dots) =  K_j(\theta)F_n^O(0,-\theta,\theta,\dots)
\end{equation*}
which are symmetric in $\theta$, i.e. the zero-momentum particle cannot
be placed between the particles $A(\theta)$ and $A(-\theta)$.

\bigskip

In the following we evaluate all $D_{nm}$ with $n+m\le
4$. We put forward that all our results are collected in
Appendix \ref{summary} together with a pictorial representation of the
individual terms.

The contributions $D_{n0}$ and $D_{0m}$ with $n,m=1\dots 4$ only
depend on one of the boundaries and they are the same as
obtained in \cite{gerard}:
\begin{equation*}
\begin{split}
    D_{10}&=\frac{g_\alpha}{2} F_1^O e^{-mx}\;,\\
  D_{20}&= \frac{1}{2}\int\frac{d\theta}{2\pi} K_\alpha(\theta) F_2^O(-\theta,\theta)
  e^{-2m\cosh\theta\ x }\;,\\
   D_{30}&= \frac{1}{2} \int\frac{d\theta}{2\pi} K_\alpha(\theta)  \frac{g_\alpha}{2} F_3^O(-\theta,\theta,0)
  e^{-m(2\cosh\theta+1)\ x }\;,\\
   D_{40}&=  \frac{1}{8}\int\frac{d\theta_1}{2\pi}\frac{d\theta_2}{2\pi} 
K_\alpha(\theta_1)K_\alpha(\theta_2) F_4^O(-\theta_1,\theta_1,-\theta_2,\theta_2)
  e^{-2m(\cosh\theta_1+\cosh\theta_2)\ x }\;,\\
  D_{01}&=\frac{g_\beta}{2}  F_1^O e^{-m(R-x)}\;,\\
  D_{02}&=  \frac{1}{2}\int\frac{d\theta}{2\pi} K_\beta(\theta) F_2^O(-\theta,\theta)
  e^{-2m\cosh\theta\ (R-x) }\;,\\
   D_{03}&= \frac{1}{2} \int\frac{d\theta}{2\pi} K_\beta(\theta)
   \frac{g_\beta}{2} F_3^O(-\theta,\theta,0)
  e^{-m(2\cosh\theta+1)\ (R-x) }\;,\\
   D_{04}&= \frac{1}{8} \int\frac{d\theta_1}{2\pi}\frac{d\theta_2}{2\pi} 
K_\beta(\theta_1)K_\beta(\theta_2) F_4^O(-\theta_1,\theta_1,-\theta_2,\theta_2)
  e^{-2m(\cosh\theta_1+\cosh\theta_2)\ (R-x)}\;. 
\end{split}
\end{equation*}
Note that these integrals remain well-defined even if there are poles in
the amplitudes
$K_j(\theta)$, because the form factors possess the appropriate number
of zeros at $\theta_i=0$ as a
consequence of the exchange axiom \eqref{eq:exchangeaxiom}. 

The first contribution to contain a divergent piece is $D_{11}$ which
is given by
\begin{equation}
 D_{11}=\lim_{L\to\infty}\Big(C_{11}-Z_1 C_{00}\Big)\;,
\end{equation}
where
\begin{equation}
  C_{11}= \frac{g_\alpha g_\beta}{4} mL\ \bra{\{0\}}O\ket{\{0\}}_L\
  e^{-mR}\;.
\end{equation}
The diagonal one-particle form factor for a generic $I$ is given by \cite{fftcsa2}
\begin{equation}
\label{diagonal1}
  \bra{\{I\}}O\ket{\{I\}}_L =\frac{F_2^{O}(i\pi+\theta,\theta)}{\rho_1(\theta)}+ \vev{O}\;.
\end{equation}
The two-particle form factor appearing in the expression above is
free of divergences and by Lorentz-invariance it does not depend on
$\theta$. In the following it will be denoted by $F_{2,s}^O$.
Specifying \eqref{diagonal1} to $I=0$ one finds
\begin{equation}
  D_{11}=\lim_{L\to\infty}\Big(C_{11}-Z_1 C_{00}\Big)= \frac{g_\alpha
    g_\beta}{4}  F_{2,s}^O
e^{-mR}\;.
\end{equation}

\subsection{Evaluation of $D_{21}$ and $D_{12}$}

We first consider 
\begin{equation}
\label{d21}
  D_{21}=\lim_{L\to\infty}\Big(C_{21}-Z_1 C_{10}\Big)
\end{equation}
with
\begin{equation}
\label{C21sum}
  C_{21}=
\frac{g_\beta}{2} 
\sum_I  \frac{F_3^O(-\theta+i\pi,\theta+i\pi,0)}{\bar\rho_1(\theta)}
 e^{-E_I x-m(R-x)} K_\alpha(\theta)\;,
\end{equation}
where it is understood that the rapidities $\theta$ are solutions of
the corresponding Bethe--\!Yang equations
\begin{equation*}
 mL\sinh\theta+\delta(\theta)+\delta(2\theta)=2\pi I\;,
\end{equation*}
and the constrained density is
\begin{equation*}
\bar\rho_1(\theta)=mL\cosh\theta+2\varphi(2\theta)\;.
\end{equation*}
Note also that we have already made use of the relations \eqref{N1N2}. 
Naively one would take the
$L\to\infty$ limit simply by replacing the summation with and integration
leading to
\begin{equation}
\frac{1}{2}  \frac{g_\beta}{2}\int\frac{d\theta}{2\pi} F_3^O(-\theta+i\pi,\theta+i\pi,0)
e^{-2m\cosh\theta  x-m(R-x)} K_\alpha(\theta)\;.
\end{equation}
However, the expression above is ill-defined at $\theta=0$ due to the
poles of $F_3$ and $K_\alpha$. Not surprisingly, this divergence gets
cancelled by $Z_1C_{10}$ as we will show shortly.

It follows from the form factor axioms \eqref{eq:exchangeaxiom} and
\eqref{eq:kinematicalaxiom} that
near $\theta_{1,2}=0$
\begin{equation}
  F_3^O(\theta_1+i\pi,\theta_2+i\pi,0)\approx 2i\ F_1^O 
\left(\frac{1}{\theta_2}-\frac{1}{\theta_1}\right)
\end{equation}
and therefore
\begin{equation}
\label{F3_polus}
  F_3^O(-\theta+i\pi,\theta+i\pi,0)\approx 4i\ F_1^O 
\frac{1}{\theta}+\dots\;.
\end{equation}

The singular behaviour of the sum \eqref{C21sum} is determined by the
states near $\theta=0$. We use the approximation of Section \ref{long-distance} to
determine the leading divergence as
\begin{equation}
   4 \frac{F_1^O}{mL} \frac{(g_\alpha)^2g_\beta}{4}
   \left(\frac{mL}{2\pi}\right)^2e^{-m(x+R)}\sum_{I} \left(\frac{1}{I}\right)^2
=
 4 \frac{F_1^O}{mL} \frac{(g_\alpha)^2g_\beta}{4}
   \left(\frac{mL}{2\pi}\right)^2e^{-m(x+R)} \frac{\pi^2}{2}\;.
\end{equation}
This coincides with $Z_1C_{10}$ therefore \eqref{d21} has a
finite $L\to\infty$ limit indeed. 

Similarly to the calculations of Section \ref{long-distance} we use
Theorem 1 to obtain the finite-left over piece in \eqref{d21}. We
apply the
substitutions 
\begin{equation*}
  f(\theta)=F_3^O(-\theta+i\pi,\theta+i\pi,0)
 e^{-2m\cosh\theta x-m(R-x)} K_\alpha(\theta)\;,\qquad\qquad
G= 2(g_\alpha)^2   F_1^O e^{-m(x+R)}\;.
\end{equation*}
It is easy to see that $f(\theta)$ is indeed a symmetric function,
therefore its only singularity near the real axis is a double pole at $\theta=0$. 
The net result is then expressed as
\begin{multline}
D_{21}= \frac{g_\beta}{4}\int\frac{d\theta}{2\pi} 
 \left(F_3^O(-\theta+i\pi,\theta+i\pi,0)K_\alpha(\theta)e^{-2m\cosh\theta  x-m(R-x)}- 
\frac{2(g_\alpha)^2 F_1^O\cosh\theta}{\sinh^2\theta}e^{-m(R+x)} \right)\\
+e^{-m(x+R)}g_\beta(g_\alpha)^2F_1^O \frac{\varphi(0)}{4}\;.
\end{multline}
A similar calculation with the roles of the two boundaries exchanged
yields
\begin{multline}
D_{12}= \frac{g_\alpha}{4}\int\frac{d\theta}{2\pi} 
\left(F_3^O(i\pi,-\theta,\theta)K_\beta(\theta)e^{-2m\cosh\theta  (R-x)-mx} - 
\frac{2(g_\beta)^2 F_1^O\cosh\theta}{\sinh^2\theta}e^{-m(2R-x)}
\right)\\
+e^{-m(2R-x)}g_\alpha(g_\beta)^2F_1^O \frac{\varphi(0)}{4}\;.
\end{multline}

\subsection{Evaluation of $D_{13}$ and $D_{31}$}

We first consider $D_{31}=\lim_{L\to\infty} (C_{31}-Z_1C_{20})$
where
\begin{multline*}
C_{31}-Z_1 C_{20}=\sum_I N_3(\theta)\sqrt{mL} \frac{g_\alpha g_\beta}{4}
  K_\alpha^*(\theta)\  \bra{\{I,-I,0\}}\mathcal{O}\ket{\{0\}}_L\
  e^{-mR-2m\cosh\theta\ x}-\\
-\frac{g_\alpha g_\beta}{4} mL e^{-mR}\sum_J N_2(\theta) \bra{\{J,-J\}} O \ket{0}_L e^{-E_I
   x} K_\alpha^*(\theta)\;.
\end{multline*}
The three-particle normalisation is given by
\begin{equation*}
  N_3(\theta)=\frac{\sqrt{\rho_3(\theta,-\theta,0)}}{\bar\rho_3(\theta)}\;.
\end{equation*}
The four-particle matrix element above includes a disconnected term due to
the zero-momentum particles which is given by \cite{fftcsa2}
\begin{equation*}
  \bra{\{I,-I,0\}}\mathcal{O}\ket{\{0\}}_L=
\frac{F_4^O(\theta+i\pi,-\theta+i\pi,i\pi,0)+mL F_2^O(\theta,-\theta)}{\sqrt{\rho_3(\theta,-\theta,0)mL}}\;,
\end{equation*}
while the connected part of $C_{31}$ can be transformed into the integral
\begin{equation}
\label{c31}
\frac{g_\alpha g_\beta}{8} e^{-mR}  \int \frac{d\theta}{2\pi}  
   K_\alpha(\theta)F_4^O(-\theta+i\pi,\theta+i\pi,i\pi,0)e^{-2m\cosh\theta\ x}\;.
\end{equation}
It follows from the form factor axioms that
$F_4^O(-\theta+i\pi,\theta+i\pi,i\pi,0)$ is well-defined  \cite{fftcsa2}
and has a
regular behaviour as a function of $\theta$. Therefore
we can apply the exchange axiom \eqref{eq:exchangeaxiom} to show that 
\begin{equation}
  \lim_{\theta\to 0} F_4^O(-\theta+i\pi,\theta+i\pi,i\pi,0) = 0\;,
\end{equation}
therefore the integral \eqref{c31} is regular even at $\theta=0$.

The divergent pieces of $D_{31}$ are given by
\begin{equation*}
  \frac{g_\alpha g_\beta}{4} e^{-mR} mL
\Big(\sum_I \frac{F_2^O(-\theta,\theta)}{\bar\rho_3(\theta)} K_\alpha(\theta)e^{-2m\cosh\theta\ x}
-\sum_J  \frac{F_2^O(-\theta,\theta)}{\bar\rho_1(\theta)} K_\alpha(\theta)e^{-2m\cosh\theta\ x}\Big)\;.
\end{equation*}
Note the summations over $I$ and $J$ run over the solutions of the
different quantisation conditions \eqref{barQ1} and \eqref{barQ3},
respectively. However, the density factors are the corresponding ones
in both sum, therefore it is allowed to replace the summation with an
integration in both terms. Doing this we make errors of order $e^{-\mu
  L}$ because both integrands are regular, therefore the above
expression vanishes in the $L\to\infty$ limit and the net result is
\begin{equation}
  D_{31}=
\frac{g_\alpha g_\beta}{8} e^{-mR}  \int \frac{d\theta}{2\pi}  
   K_\alpha(\theta)F_4^O(-\theta+i\pi,\theta+i\pi,i\pi,0)e^{-2m\cosh\theta\ x}\;,
\end{equation}
and similarly,
\begin{equation*}
D_{13}=\frac{g_\alpha g_\beta}{8} e^{-mR}  \int \frac{d\theta}{2\pi}  
   K_\beta(\theta)F_4^O(-\theta+i\pi,\theta+i\pi,i\pi,0)e^{-2m\cosh\theta\ (R-x)}\;.
\end{equation*}

\subsection{Evaluation of $D_{22}$}

Here we consider
\begin{equation}
D_{22}=\lim_{L\to\infty} \Big( C_{22}-Z_2\vev{O}-Z_1 C_{11}+(Z_1)^2\vev{O}\Big)\;.
\end{equation}
Evaluating $C_{22}$ one has to treat the diagonal and
off-diagonal matrix elements separately:
\begin{multline}
\label{C22ind}
  C_{22}
=\sum_{I\ne J} N_2(\theta_1)N_2(\theta_2) 
 \bra{\{I,-I\}}O\ket{\{J,-J\}}_L
e^{-E_I
    x-E_J(R-x)} K_\alpha(-\theta_1) K_\beta(\theta_2) +\\
+\sum_I N_2(\theta)^2 
 \bra{\{I,-I\}}O\ket{\{I,-I\}}_L
K_\alpha(-\theta) K_\beta(\theta) e^{-E_I R}\;.
\end{multline}
Here it is understood that the rapidities $\theta_{1}$ and $\theta_{2}$ solve the
quantisation condition \eqref{barQ1} with momentum quantum numbers $I$
and $J$, respectively (including the diagonal case with $\theta_1=\theta_2$ and $I=J$). 
In the off-diagonal case one has
\begin{equation}
N_2(\theta_1)N_2(\theta_2)   \bra{\{I,-I\}}O\ket{\{J,-J\}}_L=
\frac{F_4^O(\theta_1+i\pi,-\theta_1+i\pi,-\theta_2,\theta_2)}
{\bar\rho_1(\theta_1)\bar\rho_1(\theta_2)}\;.
\end{equation}
It would be desirable to convert the sum over these matrix elements
into an integral. However, one has to be careful 
because of the ambiguity of the form factors near $\theta_1=\theta_2$
and the possible poles of the integrand at $\theta_1=\theta_2=0$. As a first step we write
\begin{eqnarray}
\label{ezlennea}
\nonumber
 && \sum_{I\ne J} \frac{F_4^O(\theta_1+i\pi,-\theta_1+i\pi,-\theta_2,\theta_2)}
{\bar\rho_1(\theta_1)\bar\rho_1(\theta_2)}
 K_\alpha(-\theta_1)K_\beta(\theta_2)
e^{-2m\cosh\theta_1 x-2m\cosh\theta_2 (R-x)}=\\
\nonumber
&&=\frac{1}{4}\int\frac{d\theta_1}{2\pi}\int\frac{d\theta_2}{2\pi}
  K_\alpha(-\theta_1)K_\beta(\theta_2)
F_4^O(\theta_1+i\pi,-\theta_1+i\pi,-\theta_2,\theta_2)
e^{-2m\cosh\theta_1 x-2m\cosh\theta_2 (R-x)}\\
&&\hspace{0.5cm}-\sum_I 
\frac{\bar F_4^O(\theta+i\pi,-\theta+i\pi,-\theta,\theta)}
{\bar\rho_1(\theta)^2} K_\alpha(-\theta) K_\beta(\theta)
e^{-E_I R}+\ordo(e^{-\mu L})\;,
\end{eqnarray}
where $\bar F_4^O(\theta+i\pi,-\theta+i\pi,-\theta,\theta)$ is the
continuation of the function
$F_4^O(\theta_1+i\pi,-\theta_1+i\pi,-\theta_2,\theta_2)$ to
$\theta_1=\theta_2=\theta$.
At first sight there seems to be a double kinematic pole, however this
is not quite true. In \cite{fftcsa2} it was shown that diagonal form
factors may be assigned a finite value, although the diagonal limit will depend on
the particular evaluation method used. In the following we restate the
results of \cite{fftcsa2} which are needed to evaluate \eqref{ezlennea}.

Consider the quantity 
\begin{equation}
\label{altalanosdiag}
  F_{2n}^O(\theta_1+\eps_1+i\pi,\dots,\theta_n+\eps_n+i\pi,\theta_n,\dots,\theta_1)\;,
\end{equation}
where the singularities have been shifted off by the infinitesimal
quantities $\eps_i$. It was proven in \cite{fftcsa2} that there exists a finite limit
when all $\eps_i$ go to zero simultaneously with their ratios
fixed. Moreover, there are two special evaluation schemes which respect the
physical requirement that diagonal form factors should not depend on the order of the
rapidities. First of all,
one can consider the symmetric limit with $\eps_1=\eps_2=\dots=\eps$:
\begin{equation}
F_{2n,s}^O(\theta_1,\dots,\theta_n)\equiv
\lim_{\eps\to 0}  F_{2n}^O(\theta_1+\eps+i\pi,\dots,\theta_n+\eps+i\pi,\theta_n,\dots,\theta_1)\;.
\end{equation}
On the other hand,  one can also consider
the connected part of the diagonal form factor which is defined to be the
contribution to  \eqref{altalanosdiag} which does not contain any
singular factors of the form $\eps_i/\eps_j$ and products thereof:
\begin{equation*}
  F_{2n,c}^O(\theta_1,\dots,\theta_n)\equiv
\text{(finite part of) }  F_{2n}^O(\theta_1+\eps_1+i\pi,\dots,\theta_n+\eps_n+i\pi,\theta_n,\dots,\theta_1)\;.
\end{equation*}

The general structure of the singularities in \eqref{altalanosdiag} together
with the relation between the symmetric and connected evaluation
schemes was worked out in \cite{fftcsa2}. Here we only need the
formula relevant to the two-particle states, which states that
given two infinitesimal numbers
$\eps_1$ and $\eps_2$ one has
\begin{equation}
  F_4^O(\theta_1+\eps_1+i\pi,-\theta_2+\eps_2+i\pi,-\theta_2,\theta_1)=
F_{4,c}^O(\theta_1,\theta_2)+\left(\frac{\eps_1}{\eps_2}+\frac{\eps_2}{\eps_1}\right)
\varphi(\theta_1-\theta_2)F_{2,s}^O\;.
\end{equation}
In particular, the symmetric evaluation with $\eps_1=\eps_2$ is given
by
\begin{equation}
\label{F4sc}
 F_{4,s}^O(\theta_1,\theta_2)=  F_{4,c}^O(\theta_1,\theta_2)+
2\varphi(\theta_1-\theta_2)F_{2,s}^O\;.
\end{equation}
In the present case we need to evaluate $F_4^O(\theta_1+i\pi,-\theta_1+i\pi,-\theta_2,\theta_2)$ near
$\theta_1=\theta_2=\theta$ which
corresponds to an ''antisymmetric evaluation'' with $\eps_2=-\eps_1$:
\begin{equation}
\begin{split}
\label{antisymm}
\bar F_4^O(\theta+i\pi,-\theta+i\pi,-\theta,\theta)&\equiv \lim_{\eps\to 0} 
  F_4^O(\theta+\eps+i\pi,-\theta-\eps+i\pi,-\theta,\theta)\\
&=  F_{4,c}^O(\theta,-\theta)-2\varphi(2\theta)F_{2,s}^O\;.
\end{split}
\end{equation}
Note that the integral in \eqref{ezlennea} is well-defined even if
there are poles in the factors $K_\alpha(-\theta)K_\beta(\theta)$
because
$F_4^O(\theta_1+i\pi,-\theta_1+i\pi,-\theta_2,\theta_2)$ possesses a
double zero at
$\theta_1=\theta_2=0$ (for a proof see Appendix \ref{F4_prima}). 

In order to evaluate $C_{22}$ we also need the diagonal finite volume
form factors.
A generic diagonal four-particle matrix
element reads \cite{fftcsa2}
\begin{equation}
\label{diagonal2}
  \bra{\{I_1,I_2\}}O\ket{\{I_1,I_2\}}_L
=\frac{F_{4,s}^{O}(\theta_1,\theta_2)+(\rho_1(\theta_1)+\rho_1(\theta_2))F_{2,s}^{O}}
{\rho_2(\theta_1,\theta_2)}+ \vev{O}\;.
\end{equation}
Specifying the above formula to the present case one obtains
\begin{equation}
\label{specify}
   \bra{\{-I,I\}}O\ket{\{-I,I\}}_L
=\frac{F_{4,s}^{O}(\theta,-\theta)+2\rho_1(\theta)F_{2,s}^{O}}
{\rho_2(\theta,-\theta)}+ \vev{O}\;.
\end{equation}
Substituting the results of \eqref{ezlennea}, \eqref{antisymm} and
\eqref{specify} into \eqref{C22ind} we obtain
\begin{multline}
\label{eqreff}
\Big( C_{22}-Z_2\vev{O}-Z_1 C_{11}+(Z_1)^2\vev{O}\Big)=\\
=\frac{1}{4}\int\frac{d\theta_1}{2\pi}\int\frac{d\theta_2}{2\pi}
  K_\alpha(-\theta_1)K_\beta(\theta_2)
F_4^O(\theta_1+i\pi,-\theta_1+i\pi,-\theta_2,\theta_2)
e^{-2m\cosh\theta_1 x-2m\cosh\theta_2 (R-x)}\\
+
F_{2,s}^O \left(\sum_I \frac{2  }{\bar\rho_1(\theta)} 
K_\alpha(-\theta)K_\beta(\theta) e^{-2m\cosh\theta R}
     -\frac{(g_\alpha g_\beta)^2}{16} mL e^{-2mR}
   \right)+\ordo(e^{-\mu L})\;,
\end{multline}
where we applied the normalisation given by \eqref{N1N2}
and we also made use of the relation \eqref{F4sc} and the
identity
\begin{equation}
 \bar \rho_1(\theta)=\rho(\theta)+2\varphi(2\theta)\;.
\end{equation}

The summation in the third line of \eqref{eqreff} still contains a
divergent $\ordo(L)$ piece. However, it can be proven along the lines
of the previous subsections, that this divergence gets exactly
cancelled. One can use
Theorem 1 with the substitutions 
\begin{equation*}
  f(\theta)=2
K_\alpha(-\theta)K_\beta(\theta) e^{-2m\cosh\theta R}\;,\qquad\qquad
G=\frac{(g_\alpha g_\beta)^2}{2} e^{-2mR}
\end{equation*}
to obtain the finite left-over piece.
The net result reads
\begin{equation*}
D_{22}=  \lim_{L\to\infty} \Big( C_{22}-Z_2\vev{O}-Z_1
  C_{11}+(Z_1)^2\vev{O}\Big)=
I_{22}+J_{22}\;,
\end{equation*}
where
\begin{equation}
\label{I22eloszor}
I_{22}=\frac{1}{4}\int\frac{d\theta_1}{2\pi}\frac{d\theta_2}{2\pi}
  K_\alpha(\theta_1)K_\beta(\theta_2)
F_4^O(-\theta_1+i\pi,\theta_1+i\pi,-\theta_2,\theta_2)
e^{-2m\cosh\theta_1 x-2m\cosh\theta_2 (R-x)}  
\end{equation}
and
\begin{equation}
\label{J22-eloszor}
  J_{22}
= F_{2,s}^O \int\frac{d\theta}{2\pi} 
\Big(K_\alpha(-\theta)K_\beta(\theta) e^{-2m\cosh\theta R} -\frac{(g_\alpha
  g_\beta)^2\cosh\theta}{4\sinh^2\theta} e^{-2m R}  \Big)
+F_{2,s}^O  \frac{(g_\alpha g_\beta)^2}{8} e^{-2mR}
\varphi(0)\;.
\end{equation}

\bigskip

With this we have finished the calculation of the terms in
\eqref{complete} with $n+m\le4$. The reader can find the various terms
collected in Appendix \ref{summary} together with a pictorial
representation of them.

\subsection{Connections to other problems}

In this subsection we compare our results to those obtained in 
previous works. Also, we establish a connection to the problem
of finite temperature correlation functions.

\subsubsection{Comparison with a proposal of LeClair et al.}
\label{sec:leclair}

The first result in the literature to deal with expectation values on
a strip appeared in \cite{sachdev}. In Appendix C the authors conjecture
the general result to be
\begin{equation}
  \label{sachdev1}
  \vev{O(x)}_R=
\sum_{n=0}^\infty \sum_{\eps_i=\pm } \frac{1}{n!} I^{(n)}_{\eps_1,\dots,\eps_n}\;,
\end{equation}
where
\begin{equation}
I^{(n)}_{\eps_1,\dots,\eps_n}=\prod_{i=1}^n
 \left\{\int_{-\infty}^\infty
\frac{d\theta_i}{4\pi} g_{\eps_i}(\theta_i)
\right\}
F_{2n}^O(\theta_1,-\theta_1,\dots,\theta_n,-\theta_n)_{\eps_1,\dots,\eps_n}\;.
\end{equation}
The weight functions above are defined by
\begin{equation}
  g_+(\theta)=\frac{e^{-2mx\cosh\theta}K_\alpha(-\theta)}
{1+K_\alpha(-\theta)K_\beta(\theta)e^{-2mR\cosh\theta}}\;,
\qquad
 g_-(\theta)=\frac{e^{-2m(R-x)\cosh\theta}K_\beta(\theta)}
{1+K_\alpha(-\theta)K_\beta(\theta)e^{-2mR\cosh\theta}}\;,
\end{equation}
and the form factors are given by
\begin{equation*}
  F_{2n}^O(\theta_1,-\theta_1,\dots,\theta_n,-\theta_n)_{\eps_1,\dots,\eps_n}\equiv
F_{2n}^O(\theta_1+i\bar\sigma_1,-\theta_1+i\bar\sigma_1,\dots,\theta_n+i\bar\sigma_n,-\theta_n+i\bar\sigma_n)
\end{equation*}
with $\bar\sigma_i=\pi(1+\eps_i)/2$. It is assumed that $g_\alpha=g_\beta=0$.

It is easy to see that some of our results are reproduced by
the above series.
However, there is also a serious
discrepancy: \eqref{sachdev1} does not include the diagonal
contribution $J_{22}$ given by \eqref{J22-eloszor}. This can be proven by putting $x=R/2$ and
considering the large $R$ limit of \eqref{sachdev1}. In this case the $n$th term 
behaves as $\ordo(e^{-mnR})$ while $J_{22}$ is of
order $e^{-2mR}$. Therefore $J_{22}$ can only appear in the terms
$n\le 2$. On the other hand these contributions yield
\begin{equation*}
  \begin{split}
    I^{(1)}_+=D_{20}+\ordo(e^{-3mR})\;,\qquad   I^{(1)}_-=D_{02}+\ordo(e^{-3mR})\;,
  \end{split}
\end{equation*}
\begin{equation*}
  I^{(2)}_{++}=D_{40}+\ordo(e^{-4mR})\;,\qquad
 I^{(2)}_{--}=D_{04}+\ordo(e^{-4mR})\;,
\end{equation*}
\begin{equation*}
I^{(2)}_{+-}=I^{(2)}_{-+}=I_{22}+\ordo(e^{-4mR})\;.
\end{equation*}
All the above terms are well-defined and finite, thus it follows that $J_{22}$ is not included in \eqref{sachdev1}. 

We will demonstrate in section \ref{sec:compare} with the help of
non-perturbative numerical results that the term $J_{22}$ must be
included in the expansion, thus the series \eqref{sachdev1} can not be
complete in the general case.

\subsubsection{Relation to quench problems}

In recent papers \cite{demler-quench,davide} 
the integrable QFT approach was applied to quench problems. 
In particular, they
considered the time evolution of expectation values of local operators
after a sudden global quench which changes the Hamiltonian from $H_0$
to $H$, where $H$ (possibly also $H_0$) is considered to be integrable.
The main assumption is that the initial state of
the system (which is the ground state of $H_0$) can be expanded in the
multi-particle basis of the integrable Hamiltonian $H$ as a boundary
state. In the simplest case with only one particle type in the spectrum the corresponding expression is
\begin{equation*}
  \ket{B}=
 \exp\left(\int \frac{d\theta}{4\pi}
      K(\theta) A(-\theta)A(\theta) \right) \ket{0}\;,
\end{equation*}
where $K(\theta)$ is an arbitrary function satisfying
$K(\theta)=S(2\theta)K(-\theta)$. The time evolution of an expectation
value is then given by
\begin{equation*}
  \vev{O(0,t)}=\bra{B}e^{iHt}O(0,0)e^{-iHt}\ket{B}\;.
\end{equation*}
It is straightforward to establish a connection between the real-time
evolution of the expectation value $\vev{O(0,t)}$  
and the static (Euclidean) quantity $\vev{O(x)}^{\alpha\beta}_{R}$ considered in
this work. In fact, the quench problem above corresponds to the choice
\begin{equation*}
  K_\alpha(\theta)=K_\beta(\theta)=K(\theta)\;,
\end{equation*}
and the expectation value $\vev{O(0,t)}$   is obtained by the analytic continuation
\begin{equation*}
  R=2\tau_0\;, \qquad\qquad x=\tau_0 - it\;,
\end{equation*}
where $\tau_0$ is a constant which is analogous to the extrapolation
length introduced in the field theory approach to boundary critical phenomena
\cite{Calabrese:2006rx,davide}\footnote{Keeping a finite (non-zero) $\tau_0$
is necessary to normalise the boundary states properly,
otherwise the spectral series will not be convergent.}.

The results of \cite{demler-quench} can be reproduced by our expansion
with the analytic continuation explained above. The main difference
between \cite{demler-quench} and the present work is that
\cite{demler-quench} concerns the sine-Gordon model which is a
non-diagonal scattering theory. However, the authors calculate
contributions only from low-lying breather states which possess
diagonal scattering. Therefore our methods directly apply to the
situation in \cite{demler-quench}. Moreover, our expansion includes
higher order terms which were not considered in \cite{demler-quench}.

In the following we also compare our results to those of \cite{davide},
which considers the $t\to\infty$ limit of the expectation value $\vev{O(0,t)}$.
For this quantity the authors obtain the series expansion
\begin{equation}
\label{fioretto}
  \vev{O}=
\sum_{n=0}^\infty \frac{1}{n!}
\prod_i \left\{\int \frac{d\theta_i}{2\pi}
\frac{|G(\theta_i)|^2}{1+|G(\theta_i)|^2}\right\}
F_{2n,c}^O(\theta_1,\dots,\theta_n)\;,
\end{equation}
where the weight function is given by
\begin{equation*}
  G(\theta)=e^{-2m\tau_0\cosh\theta}K(\theta)\;.
\end{equation*}
The key idea of \cite{davide} is that in the $t\to\infty$ limit only those contributions survive which do not
depend on $t$. In our Euclidean setting this means, that the relevant
terms do not depend on $x$.
If we were able to
derive the complete spectral expansion for
$\vev{O(x)}^{\alpha\beta}_{R}$, 
then in principle it would be possible to reproduce \eqref{fioretto}
by collecting those terms which are completely independent of
$x$. However, we can already make a definite statement about the $n=1$
term in \eqref{fioretto}. It is easy to see that up to higher order
contributions this term is equal to our 
$J_{22}$ given by \eqref{J22-eloszor}, which is indeed independent of $x$.
This way we have established the first rigorous support
for the series
\eqref{fioretto}\footnote{From a
  different point of view, the presence of the term $J_{22}$ in 
  \eqref{fioretto} supports our claim that the series \eqref{sachdev1}
  (which does not include $J_{22}$) can
not be complete.}.  However, at present we can not make any
statements about the higher order terms. In
particular only a detailed study of the higher order terms in
$\vev{O(x)}^{\alpha\beta}_{R}$ can decide whether some kind of a
``dressing'' of $G(\theta)$ is needed to obtain the correct result, as it
was suggested in \cite{davide}.

\subsubsection{Relation to finite temperature correlation functions}
\label{finiteT}

There is a way to connect the vacuum expectation
values considered in this work to the evaluation of thermal correlation
functions. The relation is not physical in the sense that it can 
be established only on a formal level. However, it is certainly
worthwhile to explore this correspondence. For simplicity we suppose $g_\alpha=g_\beta=0$.
Let us define a (non-local) operator $\mathcal{B}$ by specifying its
form factors: 
\begin{multline*}
\bra{\theta_1,\theta_2,\dots,\theta_{2n-1},\theta_{2n}} 
 \mathcal{B}\ket{\theta'_1,\theta'_2,\dots,\theta'_{2m-1},\theta'_{2m}}\equiv\\
\equiv\Big(\delta(\theta_1+\theta_2) K_\alpha(-\theta_1)\Big) \dots
\Big(\delta(\theta_{2n-1}+\theta_{2n})K_\alpha(-\theta_{2n-1})\Big)\\
\times\Big(\delta(\theta'_1+\theta'_2) K_\beta(\theta'_1)\Big) \dots
\Big(\delta(\theta'_{2m-1}+\theta'_{2m})K_\beta(\theta'_{2m-1})\Big)+\Big(\dots\Big)\;.
\end{multline*}
where the dots in parentheses stand for disconnected terms which appear if two pairs
$(\theta_{2i-1},\theta_{2i})$ and $(\theta'_{2j-1},\theta'_{2j})$ coincide for some
$i=1,\dots, n$ and $j=1,\dots, m$. The form factors with an odd number of
particles on either sides are identically zero.
Note that this  definition of $\mathcal{B}$ incorporates information
about both boundaries; actually the operator itself can be visualized by gluing the two
boundaries together, thus forming a cylinder.

With this definition of $\mathcal{B}$ the vacuum expectation value
$\vev{O(x)}^{\alpha\beta}_R$ can be expressed on a formal level as
a two-point function at finite temperature $T=1/R$: 
\begin{equation*}
  \vev{O(x)}_R^{\alpha\beta}\equiv\vev{\mathcal{B}(0) O(x)}_R^{\text{periodic}}\;.
\end{equation*}
This correspondence implies, that higher order terms in the
v.e.v. ($D_{ij}$ with $i+j>4$) will present technical difficulties
which are very similar to those encountered in the evaluation of the
thermal two-point function
\cite{leclair-mussardo,saleurfiniteT,essler-2009}.  However, the exact
relation between the two objects can only be established if the
disconnected terms in the form factors of $\mathcal{B}$ are properly
defined\footnote{A naive substitution of the form factors of
  $\mathcal{B}$ without any disconnected terms results in a series
  similar to \eqref{sachdev1}. This way one misses the diagonal
  contributions like $J_{22}$. }, which is out of the scope of the
present work.

\section{Comparison with TCSA}

In this section we compare for various cases our spectral
  expansion \eqref{complete} computed in the previous section with the
  numerical results of the truncated conformal space approach
  (TCSA). We choose the scaling \LY\ model, because despite being
  one of the simplest integrable models it already features many
  properties that we have discussed so far: it contains a single
  massive particle, it has tunable integrable boundary conditions with
  different $g_j$ values and all the form factors of its elementary
  field are known. Moreover, the TCSA is quite convergent for this
  model, which together with the properties listed above makes it an
  ideal testing ground for our method.  In the first subsection we
  define the model and collect the relevant formulae which we will use
  to make the comparison in the second subsection.

\subsection{The boundary scaling \LY\ model}
\label{sec:LY}

In this section we summarise the relevant properties of the scaling
\LY\ model described as a perturbed conformal field theory. We follow
closely the treatment and notations of \cite{gerard}.

\subsubsection{The critical \LY\ model as a minimal conformal field theory}

The \LY\ model \cite{Cardy:1985yy} is the simplest example of a
non-unitary conformal field theory. It corresponds to the $M_{2,5}$
minimal model with central charge $-22/5$ and effective central charge
$2/5$. It has only two highest weight representations of the Virasoro
algebra of weights $0$ and $-1/5$, so there are only two bulk primary
fields, the identity $\One$ of weight $0$ and $\fii$ of weight
$-2/5$. 

The conformal bounary conditions of the model were classified by Cardy
in \cite{Cardy:1989ir} and the field content on each of these is given
by solving the consistency conditions written in
\cite{Cardy:1991tv,Lewellen:1991tb}. The model has only two
conformally invariant boundary conditions, denoted by $\One$ and
$\Phi$. There is only one non-trivial boundary operator
$\phi\equiv\phi_{-1/5}^{(\Phi,\Phi)}$ that can live on the $\Phi$
boundary and there are two boundary changing operators,
$\psi\equiv\phi_{-1/5}^{(\One,\Phi)}$ and
$\psi^\dagger\equiv\phi_{-1/5}^{(\Phi,\One)}$, which interpolate
between the boundary conditions $\One$ and $\Phi$. These three
boundary fields all have weight $h_\phi = h_\psi = 1/5$.

For the sake of completeness we list all the operator product
expansions (OPEs) and structure constants of the theory. The bulk OPE
is
\bea
  \fii(z,\bar z)\;\fii(w,\bar w)&=& \C\fii\fii\One \, |z-w|^{4/5} \;+\; \C\fii\fii\fii \, |z-w|^{2/5}\,\fii(w,\bar w)\;+\; \ldots\;,
\nonumber\\
\noalign{%
\vskip 1mm%
\noindent the boundary OPEs are%
\vskip 1mm}
  \phi(z)\;\phi(w)&=& \C\phi\phi\One \, |z-w|^{2/5}\;+\; \C\phi\phi\phi \,|z-w|^{1/5}\,\phi(w)\;+\; \ldots\;,\nonumber\\
  \psi(z)\;\phi(w)&=& \C\psi\phi\psi |z-w|^{1/5}\, \psi(w)\;+\; \ldots\;,\nonumber\\
  \phi(z)\;\psi^\dagger(w)&=&\!\! \C\phi{\psi^\dagger}{\!\!\psi^\dagger} |z-w|^{1/5}\,\psi^\dagger(w)\;+\; \ldots\;,\nonumber\\
  \psi(z)\;\psi^\dagger(w)&=&\C\psi{\psi^\dagger}\One|z-w|^{2/5}\,\;+\; \ldots\;,\nonumber\\
  \psi^\dagger(z)\;\psi(w)&=&\C{\psi^\dagger}\psi\One|z-w|^{2/5}\,q;+\; \C{\psi^\dagger}\psi\phi |z-w|^{1/5}\,\phi(w)\;+\;\ldots\;,\nonumber\\
\noalign{%
\vskip 1mm%
\noindent and the two bulk-boundary OPEs read%
\vskip 2mm}
 \left. \fii(z)\;\right|_\One&=& \B\One\fii\One \, |2(z-w)|^{2/5}\;+\;\ldots\;,\nonumber\\
 \left. \fii(z)\;\right|_\Phi&=& \B\Phi\fii\One \, |2(z-w)|^{2/5}\;+\; \B\Phi\fii\phi \, |2(z-w)|^{1/5}\,\phi(w)\;+\; \ldots\;.\nonumber
\eea
The structure constants can be chosen as \cite{Runkel:1998pm}
\be
{
\renewcommand{\arraystretch}{1.7}
\begin{array}{rclrcl}
\multicolumn{6}{c}{
\C\fii\fii\One ~=~ \C\phi\phi\One ~=~  -1\;,\;\;\;\;\;\;\;\;\;\C\psi{\psi^\dagger}\One = 1\;,\;\;\;\;\;\;\;\;\;\C{\psi^\dagger}\psi\One = - \fract{1 + \sqrt 5}2\;,
}
\\

\C\fii\fii\fii&= & - \left|\fract{2}{1 + \sqrt 5}\right|^{1/2} \cdot\alpha^2\;,& \B \One\fii\One&= &  -\left| \fract{2}{1 + \sqrt 5} \right|^{1/2}\;,
\\

\C{\psi^\dagger}{\psi}\phi\;=\;  \C\phi\phi\phi&= &  - \left| \fract{1 + \sqrt 5}{2}\right|^{1/2}\cdot\alpha\;,&
  \B \Phi\fii\One&= &\m  \left| \fract{1 + \sqrt 5}{2} \right|^{3/2}\;,
\\

\C\phi{\psi^\dagger}{\!\!\psi^\dagger}\;=\;  \C\psi\phi\psi&= &  - \left| \fract{2}{1 + \sqrt 5} \right|^{1/2}\cdot\alpha\;,
&
\B \Phi\fii\phi&= &\m  \left|    \fract{5 + \sqrt 5}2 \right|^{1/2}  \cdot\alpha\;,
\\

\multicolumn{6}{c}{
  \alpha ~=~  \m  \left|\fract{\Ga(1/5)\,\Ga(6/5)}{\Ga(3/5)\,\Ga(4/5)}  \right|^{1/2}\;.
}

\end{array}
}
\label{eq:ylscs}
\ee

We will need the conformal expectation value of the bulk field $\fii$
across the strip of width $R$ with various combinations of boundary
conditions: $(\One,\One)$, $(\Phi,\One)$ and $(\Phi,\Phi)$. Our
coordinates are $0\,\leq\,x\,\leq\,R$ across the strip and $y$ running
along the strip, and we normalise all our correlation functions so
that the expectation value of the identity operator is always one.

The correlation functions on the strip with boundary conditions
$\alpha$ and $\beta$ can be determined by mapping the strip to the
unit disc or the upper half plane and inserting the appropriate
boundary fields $\phi_{-1/5}^{\alpha\beta}$. Under the exponential map
$z=\exp(i\pi(x-iy)/R$ the strip can be mapped to the upper half plane
where the boundaries are mapped onto the negative and positive real
axis. The boundaries meet at the origin (and at infinity) where
boundary operators are inserted according to the boundary
conditions. Through radial quantisation the Hilbert space of the
theory is built on the representations corresponding to the boundary
operators at the origin\footnote{Thus the states on the strip, unlike
  the bulk case, are characterised by their behaviour under one copy
  of the Virasoro algebra.}. It follows that for the $(\One,\One)$
boundaries the Hilbert space consists of the $\One$ module, in the
$(\One,\Phi)$ case of the $\Phi$ module and for $(\Phi,\Phi)$ both
modules are present with the ground state corresponding to $\phi$.

If at least one of the boundaries is the $\Phi$ then the boundary
field $\phi(y)$ may exist on it. For these cases the boundary
one-point functions are
\be
\begin{array}{rcl}
  \vev{\phi(y)}_{(\Phi,\One)}=\ds  \left(\frac R\pi \right)^{1/5}\,
  \C\psi\phi\psi\;,\;\;\;\;  \vev{\phi(y)}_{(\Phi,\Phi)}= \ds  \left(\frac R\pi \right)^{1/5}\, \C\phi\phi\phi\;.
\end{array}
\label{eq:c1ptfnsb}
\ee

The one-point functions of the bulk field $\fii(x)$ on the strip
correspond in general to chiral three-point functions on the upper half
plane, and by the doubling trick they can be thought of as four-point
functions on the full complex plane. Since the representation $\phi$
has a null-vector at the second level, the correlation functions
satisfy certain differential equations, the solutions of which can be
expressed in terms of the four strip chiral block functions
\be
\begin{array}{rl}
  f_1(\vartheta)&=\ds
  \left( \frac{2 \sin\vartheta}{\cos^2\!\vartheta} \right)^{2/5}\!
  \hyp(\fract 4{10}, \fract 9{10}; \fract {11}{10}; - \tan^2\!\vartheta)
\;,
\\[3mm]
  f_2(\vartheta)&=\ds  \left( \frac{2 \sin\vartheta}{\cos^3\!\vartheta} \right)^{1/5}\!  \hyp(\fract 3{10}, \fract 8{10}; \fract {9}{10}; - \tan^2\!\vartheta)\;,
\\[3mm]
  f_3(\vartheta)&=  \,\left( {2 \sin\vartheta} \right)^{2/5}\;,
\\[2mm]
  f_4(\vartheta)&=  \,\left( {2 \sin\vartheta} \right)^{1/5}\;,
\end{array}
\label{eq:cbs}
\ee
where $\vartheta=x\pi/R$. The particular solutions are fixed by the boundary
conditions, that is the corresponding bulk-boundary OPEs:
\be
\begin{array}{rcl}
  \vev{\fii(x,y)}_{(\One,\One)}&=&\ds  \left(\frac R\pi\right)^{2/5}\, \B \One\fii\One\, f_3(\frac{\pi x}R);\;=\;\;  
  \B \One\fii\One\,  \left( \frac{2 R}{\pi}\, \sin\frac{\pi x}{ R } \right)^{2/5}\;,\;\;
\\[4mm]
  \vev{\fii(x,y)}_{(\Phi,\Phi)}&=&\ds  \left( \frac{R}{\pi} \right)^{2/5}  \,  \left(\;  \B \Phi\fii\One\; f_1( \frac{\pi x}{R} )\;+\;
  \B \Phi\fii\phi\; \C\phi\phi\phi\,f_2( \frac{\pi x}{R} )  \;\right)\;,\;\;
\\[4mm]
  \vev{\fii(x,y)}_{(\Phi,\One)}&=&\ds \left( \frac{R}{\pi}\right)^{2/5}
  \left\{  
  \begin{aligned}
   &\B \Phi\fii\One\; f_1( \frac{\pi x}{R} )\;+\;\B \Phi\fii\phi\;\C\psi\phi\psi\,f_2( \frac{\pi x}{R} )\;,
   && \qquad x\le \frac{R}2\;,\\
   &\B \One\fii\One\; f_1( \frac{\pi x}{R} )\;, &&  \qquad x>\frac{R}2\;.
  \end{aligned}
  \right.
\end{array}
\label{eq:c1ptfns}
\ee
The latter is the unique combination of the chiral blocks with the
proper asymptotic behaviour near the boundaries, for
$\vartheta\to0$ and $\vartheta \to\pi$.

Since for the $(\Phi,\Phi)$ boundary conditions both representations are
present in the Hilbert space, we will need also the non-diagonal matrix elements 
\begin{alignat}{4}
\langle\phi|\fii(x,y)|\One\rangle \;&=\;&& \left(\frac{R}\pi\right)^{1/5}\B \Phi\fii\phi \;f_4(
\frac{\pi x}{R} )\;&=\;&& \B \Phi\fii\phi \left(\frac{2 R}{\pi}\, \sin\frac{\pi x}{ R
}\right)^{1/5}\;,\\
\langle\One|\fii(x,y)|\phi\rangle \;&=\;
&-& \left(\frac{R}\pi\right)^{1/5} \B \Phi\fii\phi \;f_3(
\frac{\pi x}{R} )\;&=\; &-& \B \Phi\fii\phi \left(\frac{2 R}{\pi}\, \sin\frac{\pi x}{ R
}\right)^{2/5}\;.
\end{alignat}
Due to translational invariance in the ``time'' direction none of the
expectation values depends on $y$.

\subsubsection{The scaling \LY\ model}
\label{sec:SLY}

The scaling \LY\ (SLY) model can be defined as a bulk perturbation of
the critical \LY\ model by the term
\be
  \lambda \int \D x\int\D y \,\fii(x,y)\;.
\label{eq:bulkpert}
\ee
The resulting theory is an integrable massive scattering theory,
containing a single particle with two-particle S-matrix \cite{CM}
\be
  S(\theta)=  -(1)(2)\;,\;\;  (x) =  \frac{\sinh \left( \fract{\theta}{2} +\fract{i \pi x}{6} \right)}{\sinh \left( \fract{\theta}{2} -\fract{i \pi x}{6} \right)}\;.
\label{SMatrix}
\ee
The mass $M$ of the particle is related to the strength $\lambda$ of
the perturbation by \cite{Zb,Zg}
\be
  M = \kappa \lambda^{5/12} \;,\;\;\;\; \kappa= 2^{19/12}\sqrt{\pi}\,\ds\frac { ~\left( \Ga(3/5)\Ga(4/5) \right)^{5/12} }{5^{5/16}\Ga(2/3)\Ga(5/6)}= 2.642944\dots\;.
\label{eq:Mlambda}
\ee

The form factors of the bulk field $\fii$ were first computed in
\cite{SM1,SM2}, but following \cite{gerard} we use the conventions of
\cite{ZAM}, with the difference that for us $\fii(x)$ is a real field.
The function $F_n$ can be parametrised as
\be
  F_n(\theta_1, \ldots , \theta_n ) = H_n\,  Q_n(x_1, \ldots , x_n )\,  \prod_{i<j}^n  \frac{ f(\theta_i - \theta_j )}{x_i + x_j}\;,
\label{F-param}
\ee
where $x_i = e^{\theta_i}$.
The various terms in (\ref{F-param}) are 
\be
  f( \theta ) = \frac{ \cosh \theta -1}{\cosh \theta +1/2}\,  v(i \pi - \theta )\,  v(- i \pi + \theta )\;,
\label{f-def}
\ee
where $v$ is given by\footnote{This form with the finite product in front improves the convergence of the integral \cite{ZAM}.}
\be
\begin{array}{rcl}
    v(\theta ) &=& \ds  \prod_{n=1}^N \left[\frac{
    (\fract{\theta}{2\pi i} + n + 1/2)  
    (\fract{\theta}{2\pi i} + n - 1/6)  
    (\fract{\theta}{2\pi i} + n - 1/3)}
   {(\fract{\theta}{2\pi i} + n - 1/2)  
    (\fract{\theta}{2\pi i} + n + 1/6)  
    (\fract{\theta}{2\pi i} + n + 1/3)} 
     \right]^n       
\\[6mm]
&\times& \ds \exp\left( 2 \int_0^{\infty} \D t\, \frac{\sinh (t/2)\sinh (t/3) \sinh (t/6)}{t \sinh^2(t)}\,
(N+1 - N {\mathrm e}^{-2 t} )\,{\mathrm e}^{-2 N t + i\theta t/ \pi} \right)\;,\\[6mm]
   v(0)&=& 1.111544045...\;, 
\end{array}
\label{v-def}
\ee
and the normalisation factor is \cite{FLZZ}
\be
  H_n =  \psi  \left( -\frac{i\,3^{1/4}}{\sqrt{2} v(0)}\right)^n\;.
\label{H-def}
\ee
Here $\psi$ is the expectation value $\vev\fii$ in the bulk: 
\be
    \psi =  \frac{  - 3^{\frac 9{10}}\, \Gamma(\frac 13)^{\frac{36}5} \, M^{-\frac 2 5} }
       { (2\pi)^{\frac {14}5 }\, 5^{\frac 14 }\, \Gamma(\frac 15)\, \Gamma(\frac 25)}= (-1.239394325...\,) M^{-2/5}\;.
\label{psi}
\ee
The symmetric polynomials $Q_n(x_1 , \ldots , x_n )$ have degree $n
(n-1)/2$ and partial degree $n-1$. They can be neatly expressed as a
determinant of a matrix in symmetric polynomials
\cite{SM1,SM2,ZAM}. Here we will need only the explicit form of the
first few:
\be
  Q_0= 1  \;,\;\;  Q_1 = 1  \;,\;\;   Q_2= \sigma_1^{(2)}  \;,\;\;  Q_3= \sigma_2^{(3)} \sigma_1^{(3)}  \;,\;\;  Q_4= \sigma^{(4)}_3 \sigma^{(4)}_2 \sigma^{(4)}_1 \;,
\label{Q-ff}
\ee
where the elementary symmetric polynomials in $n$ variables
$\sigma^{(n)}_r$ are defined by
\[  \prod_{i=1}^n ( 1 + p\, x_i) = \sum_{k=0}^n p^k \sigma^{(n)}_k\;.\]

Let us turn now to the boundary conditions for the SLY model which
were analysed in detail in \cite{Us1}.  The integrable boundary
conditions are the $\One$ conformal boundary condition and the
perturbation of the conformal $\Phi$ boundary by
\be
  h \int \phi(x) \; \D x
\label{eq:boundpert}
\ee
which we will denote by $\Phi(h)$. The exact reflection factors for
these two boundary conditions are
\be
  R_{\Phi(h)}(\theta)= R_b(\theta)\;,\;\;\;\;  R_\One(\theta)= R_0(\theta)\;,
\label{eq:ss3}
\ee
where 
\be
  R_b(\theta)= \left(\fract{1}{2}\right)\left(\fract{3}{2}\right)\left(\fract{4}{2}\right)^{-1}  
\left( S(\theta+i \pi \fract{b+3}{6})  S(\theta-i \pi \fract{b+3}{6}) \right)^{-1}\;.
\label{eq:ss2}
\ee
The relation between $b$ and $h$ is given by \cite{Us1}
\be
  h(b)= - \, |\hc|\,\sin( \pi (b+1/2)/5)\;.
\ee
For numerical calculations it is more suitable to use instead the
dimensionless quantity
\[  \hhc = \hc M^{-6/5}\;,\]
which was determined in~\cite{Us3}:
\be
 \hhc=- \pi^{3/5}\,2^{4/5}\,5^{1/4}\frac{\sin\fract{2\pi} 5}{(\Ga(\fract 35) \Ga(\fract 45) )^{1/2} }  \left(   \frac{ \Ga( \fract 23 ) }{\Ga( \fract 16 )}\right)^{6/5}
= -0.68528998399118\ldots\;.
\label{eq:hexact}
\ee

Finally, let us consider the boundary-particle couplings $g_{\alpha}$
for the various boundary conditions.  We recall that these quantities are defined via the residue of the
reflection factor $R_{\alpha}(\theta)$ at the rapidity $\theta=i\pi/2$
\be
R_{\alpha}(\theta)\sim \frac{i}{2}\,\frac{(g_{\alpha})^2}{\theta-i\pi/2}\;.
\ee
In the SLY model we have
\be
  g_{\One} = -i \,2\, \sqrt{\,2\,\sqrt{~3~} - 3~}
\label{eq:gone}
\ee
for the $\One$ boundary, and 
\be
  g_\Phi(b) = \frac{ \tan((b+2)\pi/12)}{ \tan((b-2)\pi/12)}\, g_{\One}
\label{eq+gphi}
\ee
for the $\Phi(h(b))$ boundary. Notice that although the reflection
factors for the boundaries $\One$ and $\Phi(h(0))$ are identical, the
corresponding boundary-particle couplings differ by a sign.
This means that the boundary states for these two boundary conditions differ only 
in the sign of the contributions from states of odd particle number.

\subsection{The comparison}
\label{sec:compare}

\subsubsection{Expectation values in the Truncated Conformal Space Approach}
The TCSA method allows for calculating the spectrum and other physical
quantities of a perturbed conformal field theory. The approach was
developed in \cite{YZ} for the bulk scaling \LY\ model itself and
later it was generalised to systems with boundary \cite{Us1,Us3}. The
idea of the method is very simple: using the conformal field theory
techniques one can calculate in the conformal basis all the matrix
elements of the perturbing operator(s) and eventually of the perturbed
Hamiltonian. If the conformal Hilbert space is truncated at some
energy (or at some conformal level), the space of states becomes
finite dimensional and the calculation of the spectrum and of the
eigenvectors of the perturbed Hamiltonian boils down to the
diagonalisation of a finite numerical matrix. This approach can be
regarded as a close relative of the standard variational method, and
although the errors caused by the truncation can not be easily
controlled, it is generally believed, and checked of course, that they
decrease with increasing the cut. We refer the reader interested in
the details of the behaviour of TCSA to \cite{tgzserard,mienk}.

The Hamiltonian of the boundary scaling \LY\ model reads
\be
H(R;\lambda,h_l,h_r)=H_0+\lambda\int_0^R \fii(x,y)\,\D x + h_l\phi(x\text{=}0) + h_r\phi(x\text{=}R) \;,
\ee
where $H_0$ is the conformal Hamiltonian and we have allowed for
boundary perturbations at the edges of the strip (they can be present
only for $\Phi$ boundaries). The matrix elements of the Hamiltonian,
calculated by mapping the strip on the upper half plane are given by
\begin{multline}
h_{ij}
=\frac\pi{r}\Big((h_i-\frac{c}{24})\,\delta_{ij}\\
+\kappa'\left(\frac{r}\pi\right)^{12/5}(G^{-1}B)_{ij}+\chi_l\left(\frac{r}\pi\right)^{6/5}(G^{-1}B_l)_{ij} +
\chi_r\left(\frac{r}\pi\right)^{6/5}(G^{-1}B_r)_{ij}\Big)\;,
\end{multline}
where
\begin{align}
(B_l)_{ij}&=\langle i|\phi(1)|j\rangle\;,\qquad (B_r)_{ij}=\langle i|\phi(-1)|j\rangle\;,\\
(B)_{ij}&=\langle i|\int_0^\pi\D\vartheta\,\fii(e^{i\vartheta})|j\rangle\;,
\end{align}
and the conformal metric $G$ defined as $G_{ij}=\langle i|j\rangle$ is
needed because the basis vectors are not orthonormal. Here every
operator is understood to be on the upper half plane, $\vartheta=\pi x/R$
and we expressed everything in dimensionless form:
\begin{align}
r&=MR\;,\qquad \kappa'=\kappa^{12/5}=0.097048456298\dots\;, \\
\chi_l&=\sin(\pi(b_l+1/2)/5)\,\hhc\;,\qquad \chi_r=\sin(\pi(b_r+1/2)/5)\,\hhc\;,
\end{align}
implying that the energy eigenvalues are also measured in the units of
$M$.

Once the spectrum is known the expectation value of $\fii$ can be
estimated by
\be
 \vev{M^{2/5}\fii(x)}\;\sim\;  \left(\frac{r}{\pi}\right)^{2/5}\frac{ \cev{\Omega} \; \fii\,(\,\exp(i\pi \xi/r)\,)\; \vec{\Omega}}{ \veev{\Omega}{\Omega}}\;,
\label{eq:tcsavev}
\ee
where $\xi=Mx$ is the dimensionless position of the operator on the
strip and $\vec{\Omega}$ is the ground state eigenvector. If we know
the matrix elements of $\fii(\vartheta)$ on the upper half plane, the
calculation of the v.e.v. amounts to matrix-vector multiplication.

An efficient way to calculate the matrix elements is given in the
Appendix of \cite{ingothesis}.

\begin{figure}[t]
\subfigure[$(\Phi(-2),\Phi(-2))$ at $r=0.5$]{\scalebox{0.7}{\includegraphics{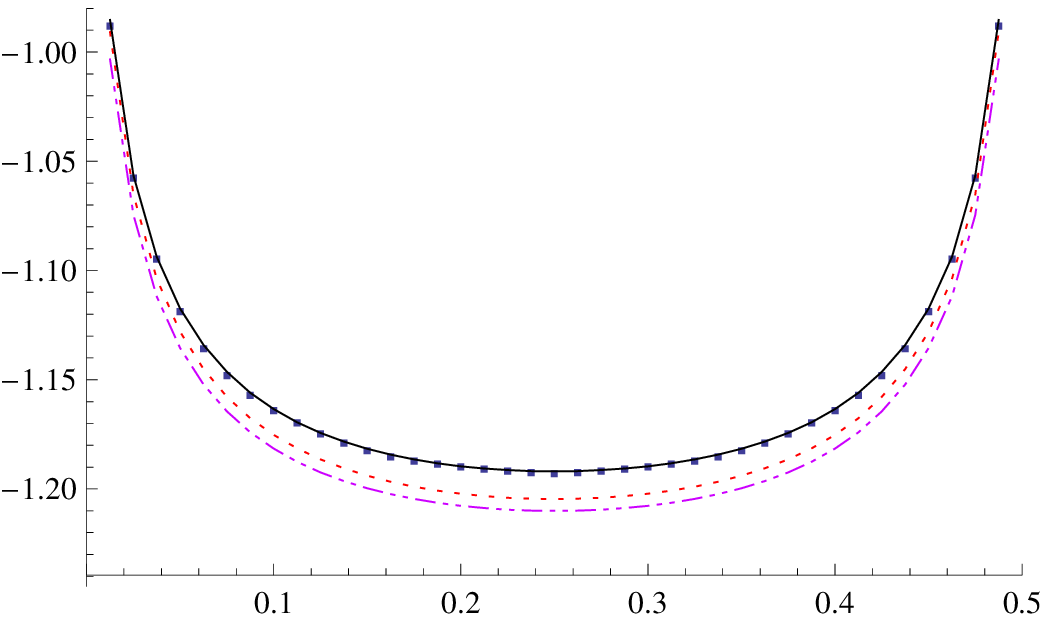}}\label{fig:2a}}
\hfill
\subfigure[$(\Phi(-1/2),\Phi(-2))$ at $r=0.5$]{\scalebox{0.75}{\includegraphics{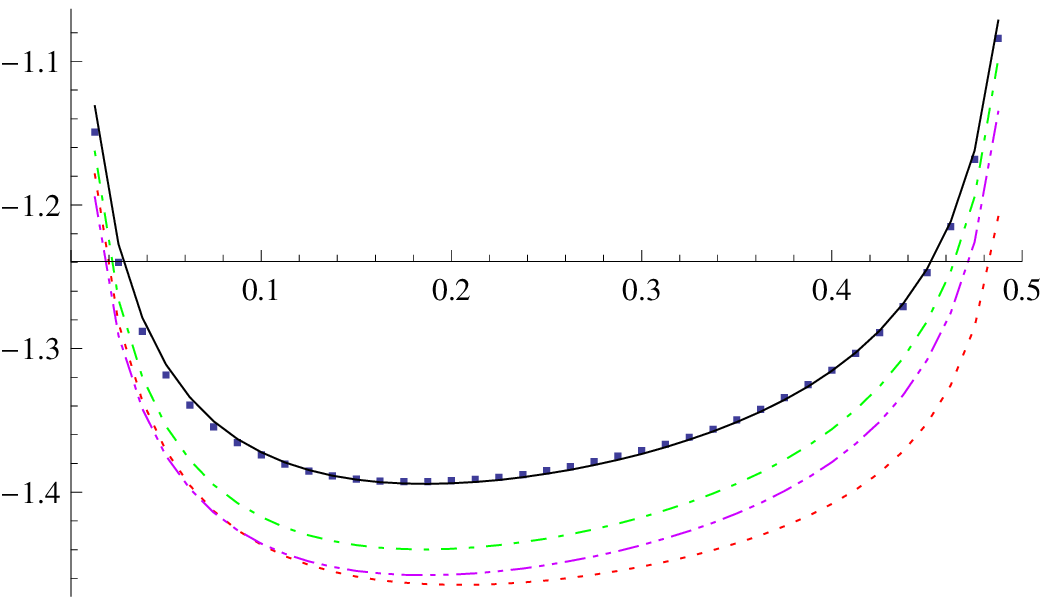}}\label{fig:2b}}
\caption{\small{$\vev{M^{2/5}\varphi(x)}$ vs. $\xi\in (0\dots r)$. The
dots are TCSA results with cut $N=10$ (48 states), the green dot-dashed and the
solid black line is our
result \eqref{eq:summary} up to the 3rd and 4th orders,
respectively. The dotted red line is \eqref{eq:2gerard} and the violet
dot-dot-dashed line is \eqref{eq:summary} without the term $J_{22}$.
The horizontal axes are positioned at the conformal $\vev{\varphi}_0$.}}
\end{figure}

\subsubsection{Comparing the form factor expansion with TCSA}

Let us recall our main result that we will compare with the TCSA data: 
\begin{equation}
 \vev{O(x)}_{R}^{\alpha\beta}=\vev{O}+\sum_{i+j\le   4} D_{ij}+\dots\;.
\label{eq:summary}
\end{equation}
In the following we will refer to terms $D_{ij}$ with $i+j=n$ as $n$th
order terms.

The work \cite{gerard} addressed the calculation of the vacuum
expectation value in large volume when the operator is close to one of
the boundaries. With this setup they could neglect the effect of the
other boundary and they worked up to the third order, so their formula in our notations is 
\begin{equation}
\vev{O}+D_{10}+D_{20}+D_{30}\;.
\label{eq:gerard}
\end{equation}
One can try to improve this result for finite strips by considering
also the analogous terms coming from the other boundary:
\begin{equation}
\vev{O}+\sum_{i=1}^3 (D_{i0}+D_{0i})\;.
\label{eq:2gerard}
\end{equation}
In order to show that our formula means a great improvement we will compare
this expression with our result \eqref{eq:summary}.

Figure \ref{fig:2a} shows the TCSA data with cut $N=10$ (48 states)
together with our results for $\vev{M^{2/5}\varphi(\xi)}$ for the boundary
conditions $(\Phi(b=-2),\Phi(b=-2))$. In this case
$g_\alpha=g_\beta=0$ and only the even orders are non-zero. In
particular, in this special case formula \eqref{eq:2gerard} coincides
with our second order result. In Figure \ref{fig:2b} we plot the case
of the boundaries $(\Phi(b=-1/2),\Phi(b=-2))$. The TCSA has already
converged, since changing the cut from $N=9$ (38 states) to $N=10$ (48
states) the maximum
change in the v.e.v. is $\approx0.0005$. The plots clearly show
that our formula gives much better results than \eqref{eq:2gerard}
throughout the strip. We also plotted our result without the term
$J_{22}$ (violet dot-dot-dashed lines) in order to show that this non-trivial
term is indeed present in the expansion (see the discussion in section
\ref{sec:leclair}).

\begin{figure}[t]
\subfigure[$(\One,\Phi(0))$ at
  $r=3$]{\scalebox{0.75}{\includegraphics{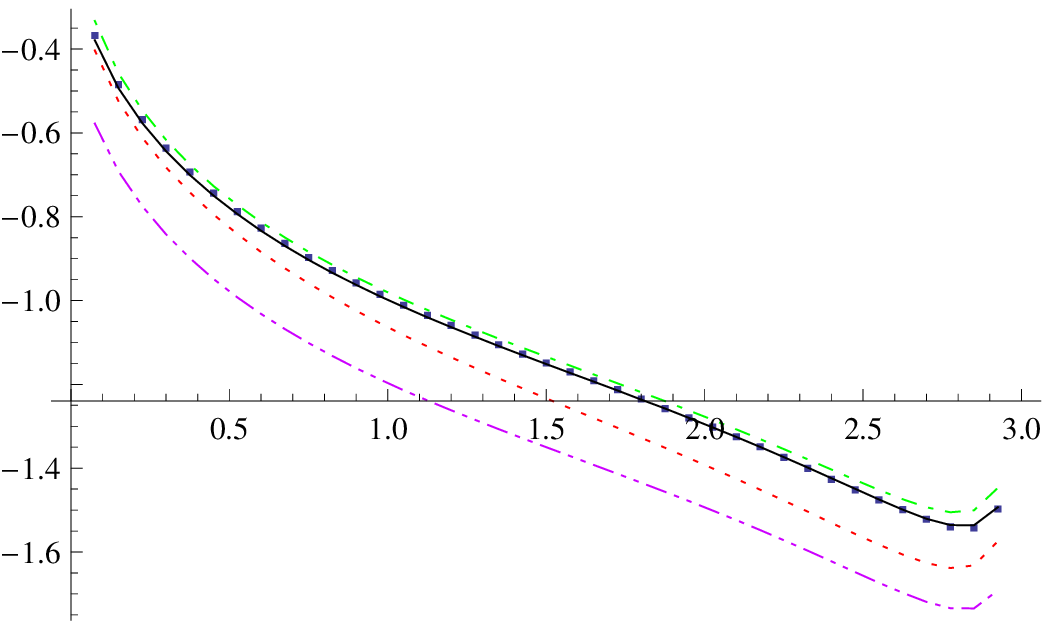}}}
\hfill
\subfigure[$(\Phi(-1/2),\Phi(-1/2))$ at $r=2$]{\scalebox{0.75}{\includegraphics{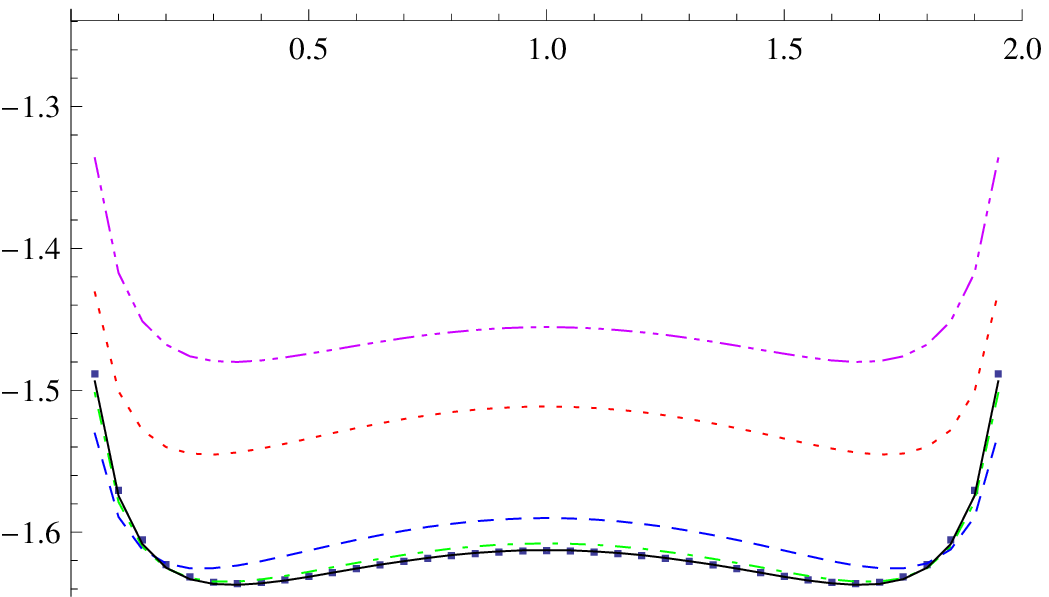}}\label{fig:0505}}
\caption{\small{$\vev{M^{2/5}\varphi(x)}$ vs. $\xi\in (0\dots r)$. The
dots are TCSA results with cut $N=12$ (45 states) in {\bf (a)} and
$N=10$ (48 states) in {\bf
  (b)}, the blue dashed, the green dot-dashed and the solid black
line is our result \eqref{eq:summary} up to the 2nd, 3rd and 4th orders,
respectively. The dotted red line is \eqref{eq:2gerard} and the violet
dot-dot-dashed line is \eqref{eq:summary} with the sign of the term $D_{11}$ flipped. The
horizontal axes are positioned at the conformal $\vev{\varphi}_0$.}}
\label{fig:szepek}
\end{figure}

Figure \ref{fig:szepek} shows the v.e.v.\ for other pairs of boundary
conditions. For the second pair changing the TCSA cut from $N=9$ to $N=10$
resuls in a change of order $10^{-5}$ in the v.e.v. For the
$(\One,\Phi(0))$ boundaries raising the cut from $N=8$ (18 states) to
$N=12$ (45 states) the change in the v.e.v. is only $\approx0.05$. 

In order to show that the conventions for the complex conjugation of
the $g$ factors are correct we plot also the result with the sign of
the term $D_{11}$ reversed (violet dashed lines). In Figure
\ref{fig:0505} we show also the second order result to show the
convergence of the form factor expansion. It is clearly seen in both
plots that adding the fourth order terms hardly changes the result,
thus in these cases the form factor series is almost saturated by its
first four orders.

\begin{figure}[t]
\subfigure[$(\One,\One)$ at $r=3$]{\scalebox{0.75}{\includegraphics{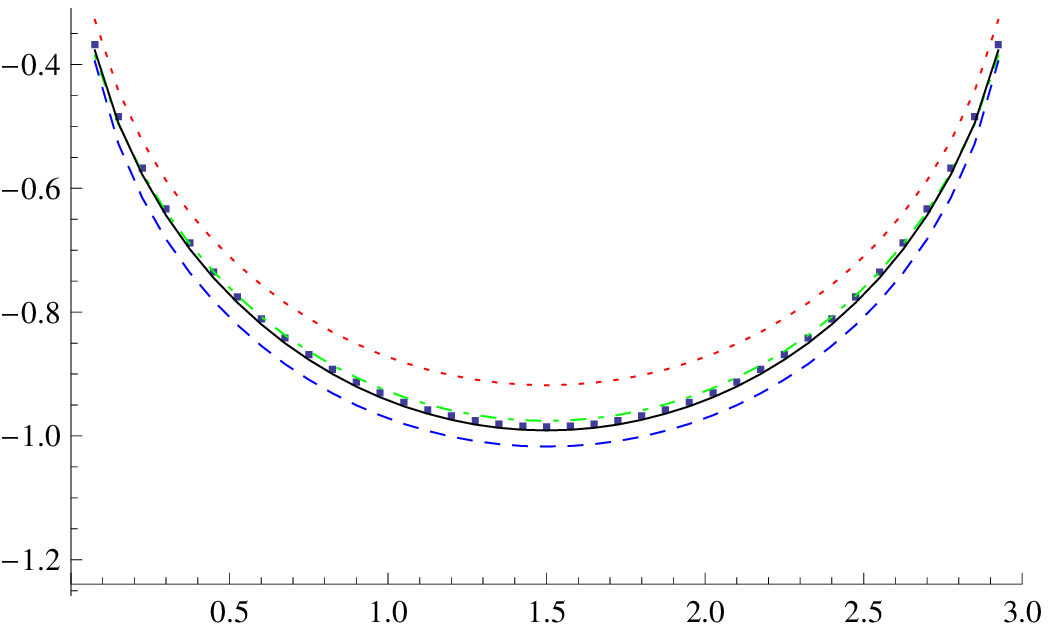}}}
\hfill
\subfigure[$(\One,\One)$ at $r=2$]{\scalebox{0.75}{\includegraphics{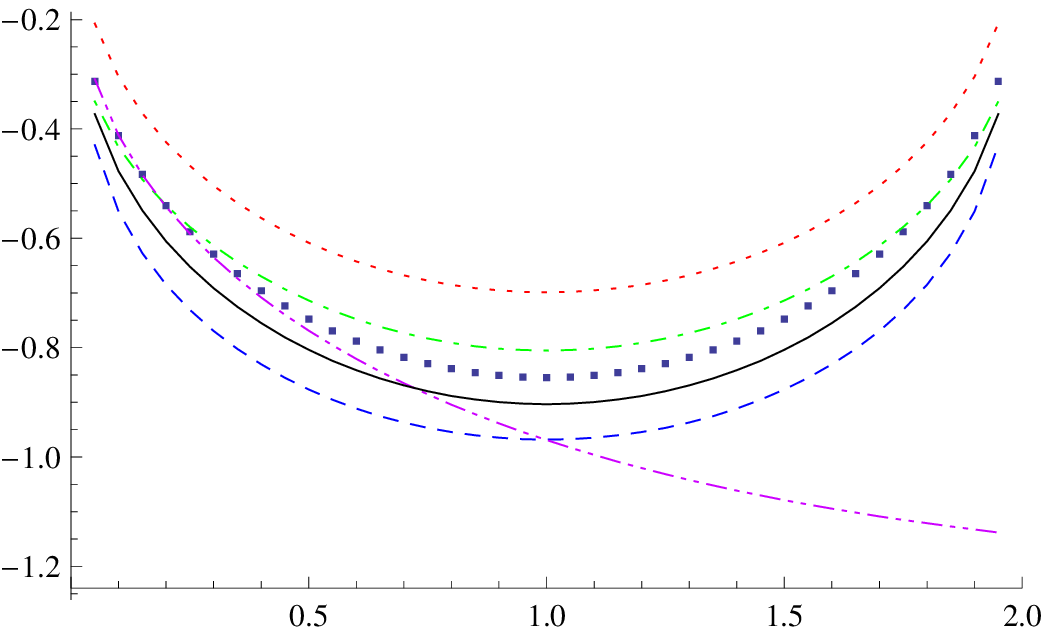}}}
\caption{\small{$\vev{M^{2/5}\varphi(x)}$ vs $\xi\in (0\dots r)$. The
dots are TCSA results with cut $N=12$ (29 states), the blue dashed, green
dot-dashed and the solid black line is our
result \eqref{eq:summary} up to the 2nd, 3rd and 4th orders, respectively. The dotted red
line is \eqref{eq:2gerard} and the asymmetric violet dot-dot-dashed line in {\bf (b)} is \eqref{eq:gerard}. The
horizontal axes are positioned at the conformal $\vev{\varphi}_0$.}}
\label{fig:id}
\end{figure}

In Figure \ref{fig:id} we plot the one-point function for the
$(\One,\One)$ boundaries for different strip widths. The TCSA is very
convergent also in this case: a change in the cut from $N=8$ (12
states) to $N=12$ (29 states) causes a shift of order
$10^{-5}$ in the v.e.v. The convergence of the form factor expansion is again
obvious, however, for $r=2$ our formula starts to deviate from the
TCSA values, as a clear sign of the need for higher order terms. Our
result is closer to the real values than the formula
\eqref{eq:2gerard}, but what is surprising is that the result of
\cite{gerard}, expression \eqref{eq:gerard}, performs very well in the
vicinity of the boundary. This can be understood by looking at the
leading behaviour of the expectation value near the different
boundaries \cite{gerard}:
\begin{subequations}
\begin{align}
\vev{M^{2/5}\varphi(x)}_{\Phi(0)}
&= 
  (2\xi)^{1/5}\,\B\Phi\varphi\phi\,\vev{\phi}_{\Phi(0)}+(2\xi)^{2/5}\,\B\Phi\varphi\One\,\vev{\One}_{\Phi(0)}+\ordo(\xi^{12/5})\;,
\label{eq:exp1}
\\
\vev{M^{2/5}\varphi(x)}_{\One}
&= 
  (2\xi)^{2/5}\,\B\One\varphi\One\,\vev{\One}_{\One}+\ordo(\xi^{12/5})\;.
\label{eq:exp2}
\end{align}
\end{subequations}
We have to check what happens when $x$ is fixed and the volume $R$ is
changed. We know that the corrections to \eqref{eq:gerard} are of the
form $\exp(-M R)$, thus rescaling $R$ is equivalent to a
rescaling of $M$. Equation \eqref{eq:exp2} remains unchanged under
such a rescaling in the leading order (and the next term is two orders
higher in $\xi$). In other words, the $x$-dependence of the
v.e.v. agrees with the behaviour under a conformal rescaling, which means that for the $\One$ boundary the expectation value will not
change much when the width of the strip changes. Obviously this is not true for \eqref{eq:exp1}
\footnote{We would like to
  thank G\'erard Watts for discussing this question.}.

\begin{figure}[t]
\subfigure[linear scale]{\scalebox{0.75}{\includegraphics{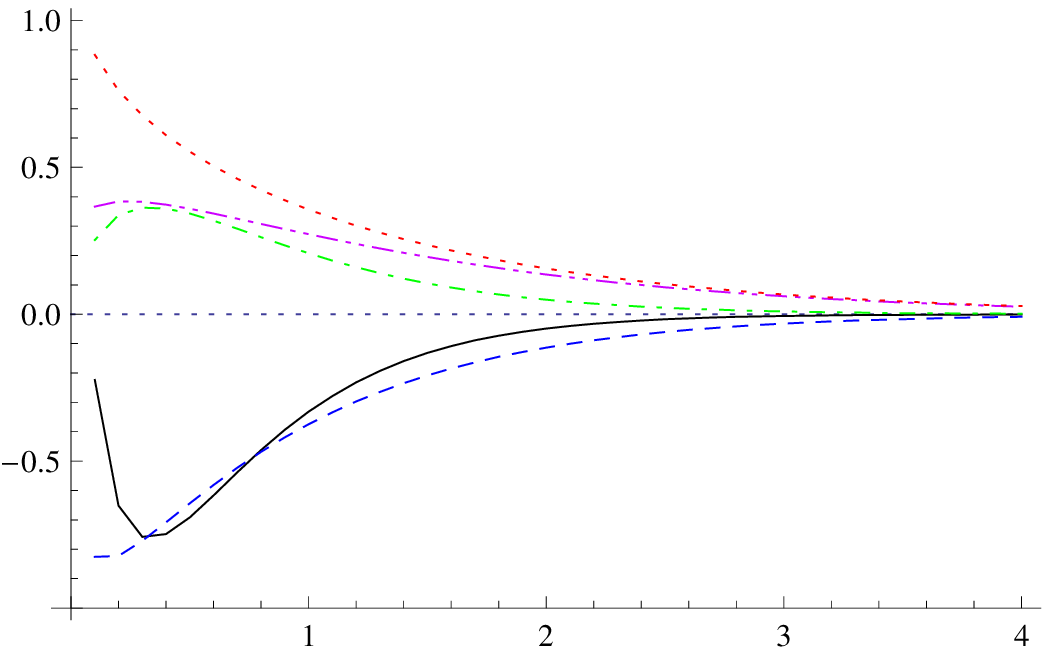}}}
\hfill
\subfigure[log scale]{\scalebox{0.75}{\includegraphics{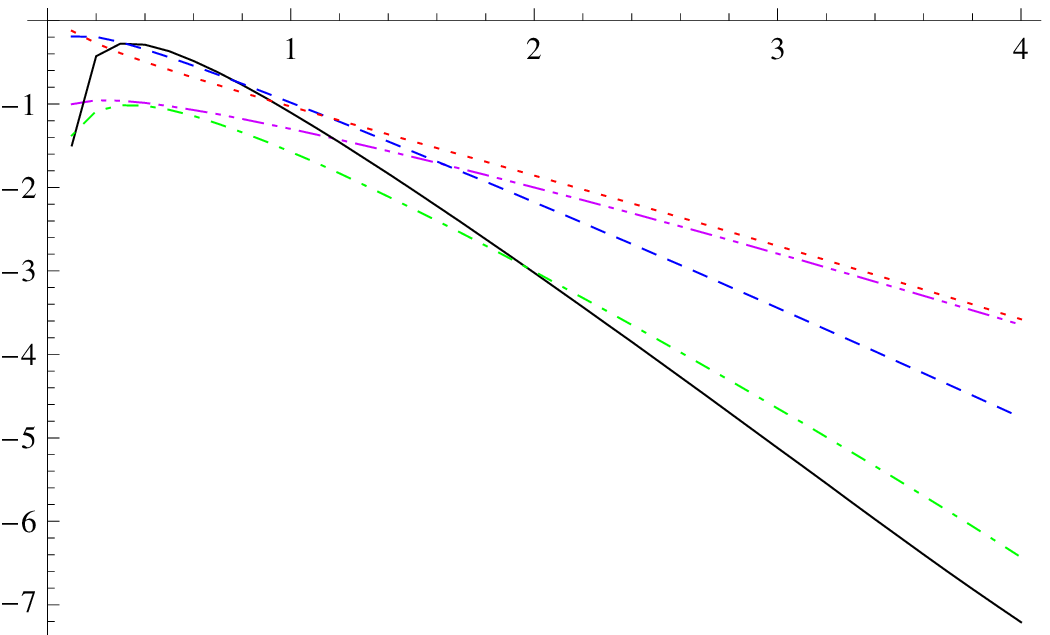}}}
\caption{$\vev{M^{2/5}\varphi(x=\frac{R}2)}-\vev{M^{2/5}\varphi(\frac{R}2)}_{\text{TCSA}}$
  for b.c. $(\One,\One)$ with TCSA cut $N=12$ (29 states) as a function of
  $r$. The violet dot-dot-dashed, the blue dashed, the green
  dot-dashed and the black solid line is our result \eqref{eq:summary}
  up to the 1st, 2nd, 3rd and 4th order, respectively. The red dotted
  line is formula \eqref{eq:2gerard}. For $r\gtrsim2$ the errors decrease
  exponentially with $r$.}
\label{fig:idmid}
\end{figure}

Since our truncation of the form factor series is truly consistent
only at $x=R/2$, in Figure \ref{fig:idmid} we plot the deviation of
the v.e.v. from the TCSA values at the middle of the strip for
different volumes $R$. It is clearly seen on the logarithmic plot
  that for large enough volume ($r\gtrsim2$) the errors decrease
  exponentially with the volume and the exponents for the error of the
  consecutive orders are approximately even-spaced. In particular, the
  sum of all four orders has the smallest and most rapidly decreasing
  error. A similar behaviour can be observed for other boundary conditions.

\section{Conclusions}

In this paper we considered vacuum expectation values of local
operators in the presence of two integrable boundaries. We developed a
consistent finite volume regularisation scheme to handle the various
types of disconnected terms. The method presented in this work can be
considered as a generalisation of the approach of \cite{fftcsa2} which
dealt with expectation values with periodic boundary conditions.  The
key ingredients of our approach are the proper normalisation of the
boundary state in a finite volume and the knowledge of the finite
volume form factors.

Apart from the task of evaluating disconnected terms of the form
factors the two-boundary setting poses additional difficulties for
$g_\alpha g_\beta\ne 0$. In this case the boundary states include
zero-momentum particles corresponding to the poles of the amplitudes
$K_j(\theta)$ at $\theta=0$. We showed that our finite volume
evaluation scheme automatically provides a regularisation of the
resulting singularities. We developed a new method to extract the
finite parts (see Theorem 1 in section \ref{thmsec}) which can be
applied also in other cases, e.g.\ two-point functions at finite
temperature.

As a by-product of our formalism we were able to extract the second
order terms of the ground state energy on the strip (subsection
\ref{long-distance}). This way we confirmed the boundary Thermodynamic
Bethe Ansatz (BTBA) equations and we showed that they yield the
correct second order terms even in the presence of zero-momentum
particles ($g_\alpha g_\beta\ne 0$), for which no rigorous derivation
of the BTBA is known. Also, we gave further independent support for
the normalisation condition $\bar g_j=g_j/2$, which was proven by
other means in \cite{Bajnok:2004tq,Bajnok:2006dn}.

It is possible to compute higher order terms of the vacuum expectation
value $\vev{O(x)}^{\alpha\beta}_R$ using the methods laid out in this
work; in particular the finite volume normalisation of higher
multi-particle contributions to the boundary state follows in a
straightforward manner.  However, the evaluation of the $L\to\infty$
limit poses technical difficulties already at the next orders
($D_{ij}$ with $i+j>4$). There will appear new types of singularities
which are very similar to those encountered in the evaluation of
thermal correlation functions (see subsection \ref{finiteT}). These
difficulties call for new summation schemes to separate the divergent
and finite parts of the relevant finite volume series. Alternatively,
it would be interesting to develop a general infinite volume
regularisation scheme along the lines of
\cite{essler-2009,esslerIsing}, which could be compared to the results
of the present approach.

There is plenty of room for future extensions of this work. First of
all, one can consider expectation values with respect to a boundary
excited state. In this case one has to develop an appropriate
modification of the finite volume boundary state formalism.  A further
interesting problem is the evaluation of (space and time dependent)
correlation functions on the strip, which can be performed along the
lines of the present paper.  Future work is needed to generalise our
methods to non-diagonal scattering theories which are relevant to a
number of condensed matter problems \cite{esslerfiniteT}.  Also, it
needs to be clarified whether some of the present results also apply
to massless scattering theories. In the absence of a mass gap the
convergence of the spectral series is not guaranteed anymore and one
does not have control over residual finite size corrections to form
factors. However, in some cases it is possible to apply a modified
version of the form factor program to massless theories
\cite{Delfino:1994ea,Cabra:1997rc}. The study of a finite volume
regularisation scheme for massless theories is left as an open
problem.

\vspace{0.7cm}

{\bf Acknowledgements} 
We would like to thank G\'abor Takacs and G\'erard Watts for useful
discussions. The early stages of this work carried out while
B. P. was a research assistant in the Theoretical Physics Group of the
Hungarian Academy of Sciences. Also, he was partially supported by the
grant OTKA K75172 of the HAS. M. K. was supported 
by the grants INSTANS (from ESF) and 2007JHLPEZ
(from MIUR).

\vspace{1cm}

{\bf\Large Appendix}

\appendix

\section{Regularisation of the divergent sums}

\label{Prooff}

For a given $L$ let us denote by $\theta_I$ the
solutions of the quantisation condition
\begin{equation}
\label{eq:Qkv}
 Q(\theta)= mL\sinh\theta+\delta(2\theta)=2\pi I\;,\qquad I\in \mathbb{N}+\frac{1}{2}\;.
\end{equation}
where $\delta(\theta)$ is the elastic phase shift defined by
\begin{equation*}
  S(\theta)=-e^{i\delta(\theta)},\qquad\qquad
\delta(-\theta)=-\delta(\theta)\;.
\end{equation*}
We also introduce the density of states
\begin{equation*}
  \bar\rho_1(\theta)=\frac{dQ(\theta)}{d\theta}=mL\cosh\theta+2\varphi(2\theta)\;,
\end{equation*}
where $\varphi(\theta)=\delta'(\theta)$.

{\it{\bf Theorem 1:}
Let $f(\theta)$ be a
   symmetric function which apart from a double pole at $\theta=0$ is
   analytic in a neighbourhood of the real axis:
\begin{equation*}
  f(\theta)\approx \frac{G}{\theta^2} \quad \text{as}\quad
    \theta\to 0\;. 
\end{equation*}
Then the expression
\begin{equation*}
  S(L)=\left(\sum_I \frac{f(\theta_I)}{\bar\rho_1(\theta_I)}\right)-\frac{G}{8}mL
\end{equation*}
has a regular behaviour at large $L$ with the $L\to\infty$ limit given by
\begin{equation*}
  \lim_{L\to\infty} S(L)=I_f+K_f\;,
\end{equation*}
where
\begin{equation*}
\begin{split}
  I_f= \int_{-\infty}^\infty \frac{d\theta}{4\pi} \left(
f(\theta)-G\frac{\cosh\theta}{\sinh^2\theta}
\right)\qquad\text{and}\qquad
K_f=\frac{G}{4} \varphi(0)\;.
\end{split}
\end{equation*}
}

{\bf Proof}:

We express the sum 
 in the form of a complex integral
for a finite $L$ and perform the $L\to\infty$ limit afterwards. First
of all, the
summation over the rapidities can be replaced by a sum over contour
integrals around the solutions of $e^{iQ}+1=0$:
\begin{equation}
  \label{kont1}
- \frac{1}{2} \sum \oint \frac{1}{2\pi}\frac{f(\theta) }{e^{iQ}+1}\;,
\end{equation}
where now the summation is over $I\in \mathbb{Z}+1/2$.
The integration contours can be
transformed into two distinct curves:
\begin{itemize}
\item the first starting from $\theta=\infty + i\eps$ running to $\eps+i\eps$,
  crossing the real axis between 0 and the first solution of
  \eqref{eq:Qkv}, then running from $\theta=\eps-i\eps$ to
  $\theta=\infty-i\eps$;
\item and a similar curve around the negative real axis, with the same
  counter-clockwise orientation.
\end{itemize}
These two curves can be joined to form a single contour encircling the whole
real axis. In
doing this, one picks up the residue at $\theta=0$, which is given by
\begin{equation*}
\frac{1}{2}\mathop{\text{Res}}_{\theta=0} 
\frac{f(\theta) }{e^{iQ(\theta)}+1}
=
-i\frac{G}{8} \bar\rho_1(0)=
-i\frac{G}{8} \left(mL+2\varphi(0)\right)\;.
\end{equation*}
Therefore the $\ordo(L)$ terms cancel and we obtain
\begin{equation*}
\begin{split}
S(L)= \left(\int_{-\infty+i\eps}^{\infty+i\eps}-\int_{-\infty-i\eps}^{\infty-i\eps}\right)
\frac{1}{4\pi}
\frac{f(\theta)}{e^{iQ}+1}
+ \frac{G}{4}\varphi(0)\;.
\end{split}
\end{equation*}

Note that in the two integrals the $L$ dependence is only contained in
$Q(\theta)$. We may now perform the $L\to\infty$ limit for a fixed
$\eps>0$. To do this first of all note that the integrand has become
bounded (we stay away from the real axis) therefore one may exchange
the limit $L\to\infty$ with the integration. Also, observe that
\begin{equation*}
\lim_{L\to\infty} \frac{1}{e^{iQ(\theta+i\eps)}+1}=1\;,\qquad\qquad
\lim_{L\to\infty} \frac{1}{e^{iQ(\theta-i\eps)}+1}=0\;.
\end{equation*}
Putting everything together one obtains 
\begin{equation}
  \label{eq:majdnem_vege2}
 \int_{-\infty+i\eps}^{\infty+i\eps}
\frac{1}{4\pi}
f(\theta)
+ 
\frac{G}{4} \varphi(0)\;.
\end{equation}
The integral above can be pulled back to the real axis after an
appropriate regularisation. We use the identity
\begin{equation}
  \label{eq:nulla}
  0=\int_{-\infty+i\eps}^{\infty+i\eps}
\frac{1}{4\pi} G
 \frac{\cosh\theta}{\sinh^2\theta}
\end{equation}
which can be proven by deforming the contour to
$\text{Im}\theta=i\pi/2$. Moreover, the function above has exactly the
same singularity structure around $\theta=0$ as the integrand in
\eqref{eq:majdnem_vege2}. One may therefore subtract \eqref{eq:nulla}
from \eqref{eq:majdnem_vege2} and pull back the integration contour to
the real axis. The final result is thus given by
\begin{equation*}
  \lim_{L\to\infty} S(L)=
 \int_{-\infty}^\infty \frac{d\theta}{4\pi} \left(
f(\theta)-G\frac{\cosh\theta}{\sinh^2\theta}
\right)+\frac{G}{4} \varphi(0)\;.
\end{equation*}

\section{Properties of the four-particle form factor}

\label{F4_prima}

In this appendix we consider the behaviour of the function
\begin{equation}
\label{F4fu}
  F_4^O(\theta_1+i\pi,-\theta_1+i\pi,-\theta_2,\theta_2)
\end{equation}
near $\theta_1,\theta_2=0$. It follows from the form factor axioms
that for infinitesimal $\theta_{1,2}$ the function above is
antisymmetric in both variables and disappears whenever one of the
rapidities goes to zero whith the other one being kept fixed.
Therefore the only allowed singularity is of the form
\begin{equation*}
  \frac{1}{\theta_1\theta_2}\;.
\end{equation*}
However, it is easy to see that this pole does not appear.
Setting
\begin{equation}
\theta_1=\theta+\eps\;,  \qquad\qquad \theta_2=\theta\;,
\end{equation}
and taking $\eps\to 0$ results in \cite{fftcsa2}
\begin{equation}
\begin{split}
&\lim_{\eps\to 0}
F_4^O(\theta+\eps+i\pi,-\theta-\eps+i\pi,-\theta,\theta)=
\big(F_{4c}^O(\theta,-\theta)-2\varphi(2\theta)F_{2c}^O\big)\;,
\end{split}
\label{K12_szingub}
\end{equation}
which is well-defined for every $\theta$. Therefore a pole of
$\frac{1}{\theta_1\theta_2}$ 
cannot appear and it follows that
\begin{equation*}
\lim_{\theta_1,\theta_2\to 0} F_4^O(\theta_1+i\pi,-\theta_1+i\pi,-\theta_2,\theta_2)=
\lim_{\theta\to 0} \big( F_{4c}^O(\theta,-\theta)-2\varphi(2\theta)F_{2c}^O\big)= 0\;.
\end{equation*}
Moreover, 
the first term
of its Taylor expansion is proportional to $\theta_1\theta_2$,
as required in \eqref{I22eloszor} to cancel the double pole of
$K_\alpha(-\theta)K_\beta(\theta)$. 

\newpage

\section{Summary of our results}

\label{summary}

\begin{figure}[p]
  \centering

 \subfigure[$D_{10}$]{\includegraphics[scale=0.4]{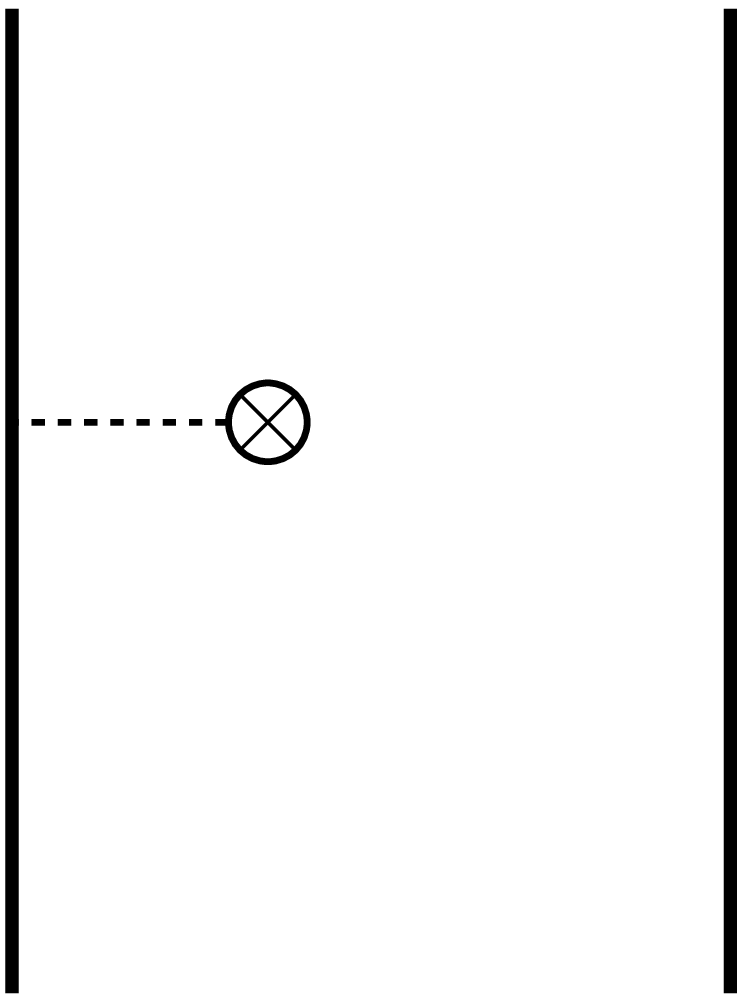}\label{F1}}  
\hspace{3.5cm}
  \subfigure[$D_{20}$]{\includegraphics[scale=0.4]{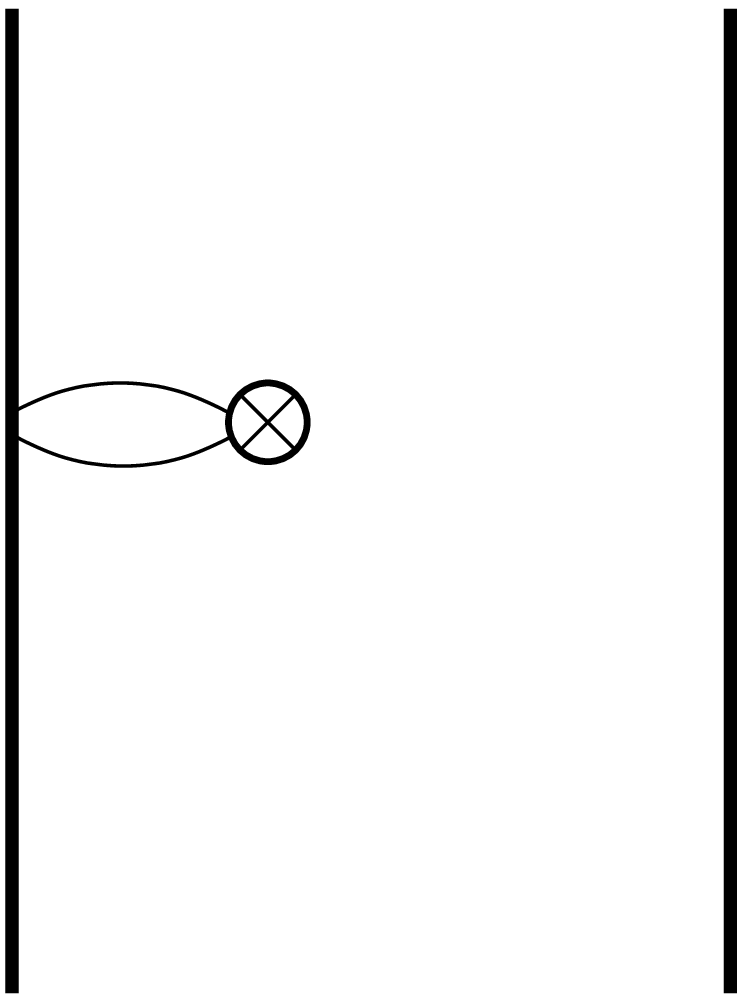}\label{I1}}  

\subfigure[$D_{30}$]{\includegraphics[scale=0.4]{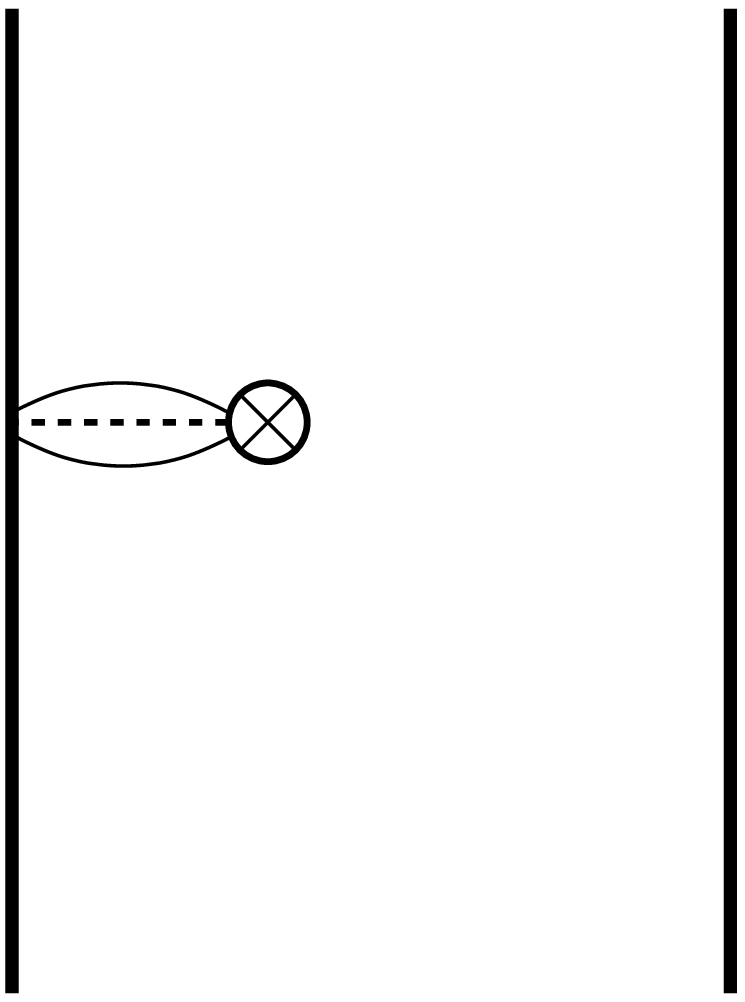}\label{D30}}
\hspace{3.5cm}
 \subfigure[$D_{40}$]{\includegraphics[scale=0.4]{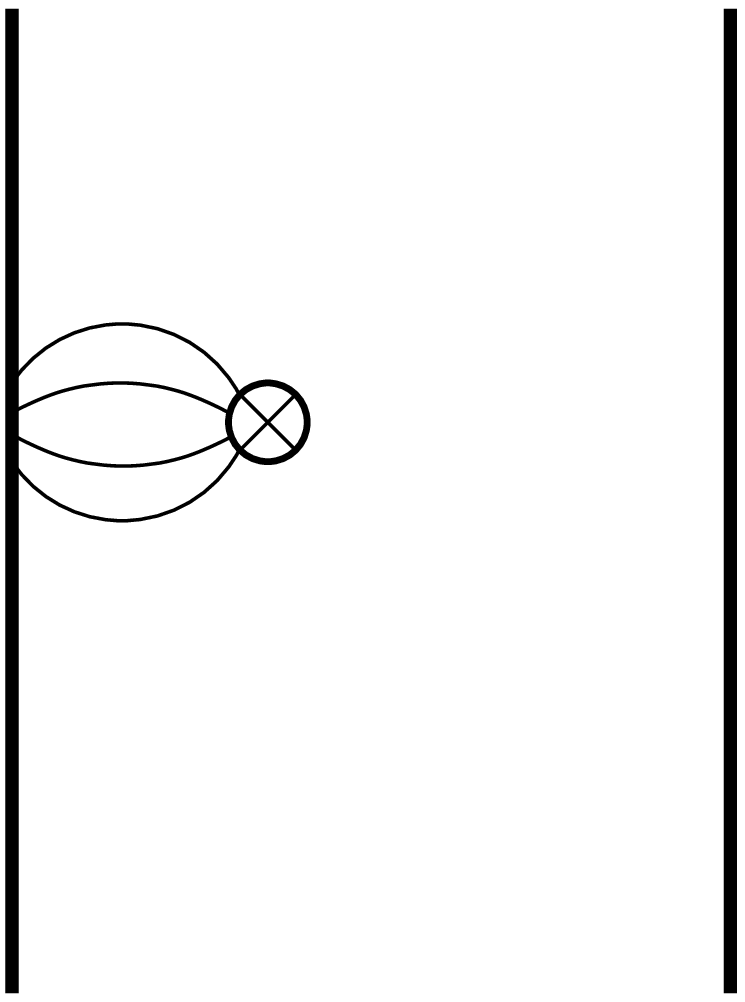}\label{D40}} 

\subfigure[$D_{11}$]{\includegraphics[scale=0.4]{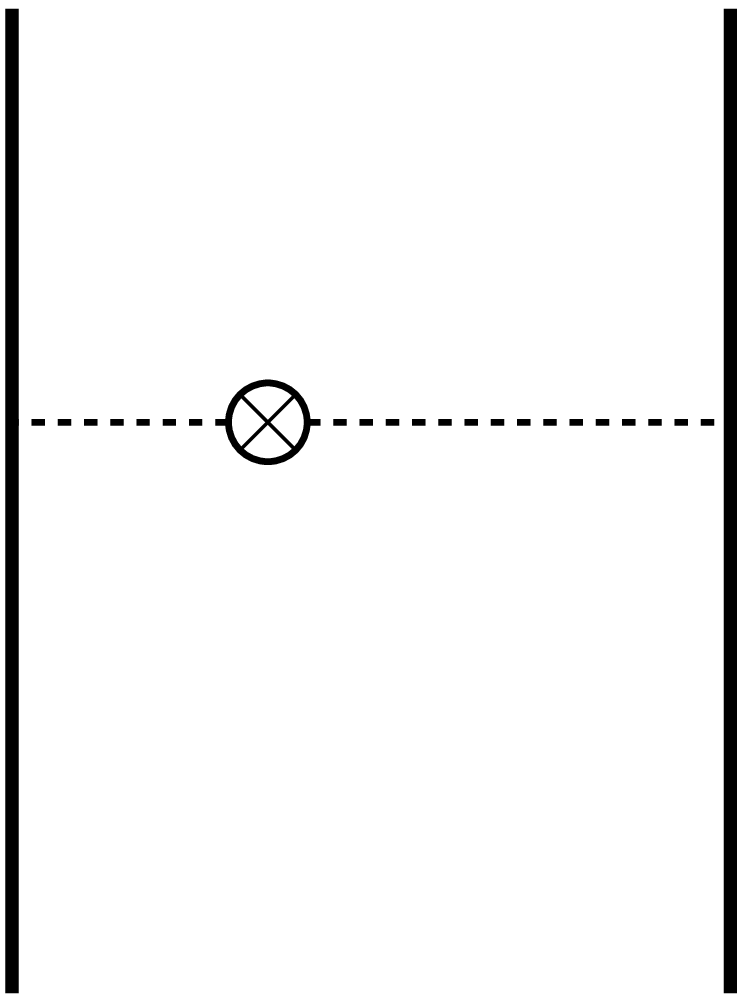}\label{F12}}
\hspace{2cm}
 \subfigure[$D_{31}$]{\includegraphics[scale=0.4]{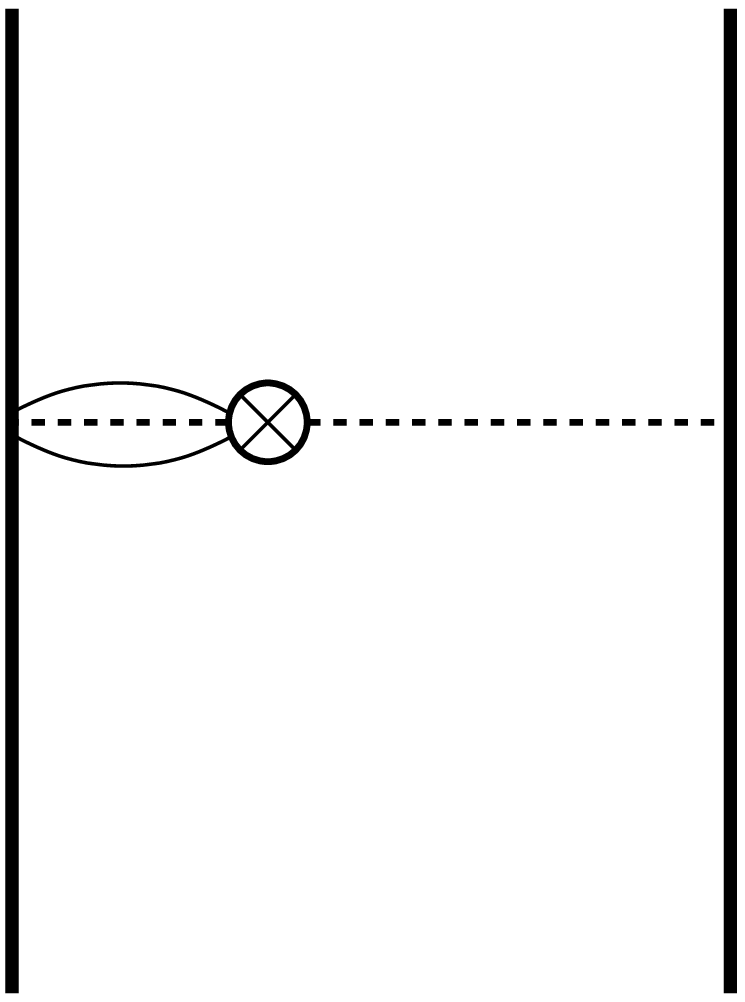}\label{P1}}  
\hspace{2cm}
\subfigure[$I_{22}$]{\includegraphics[scale=0.4]{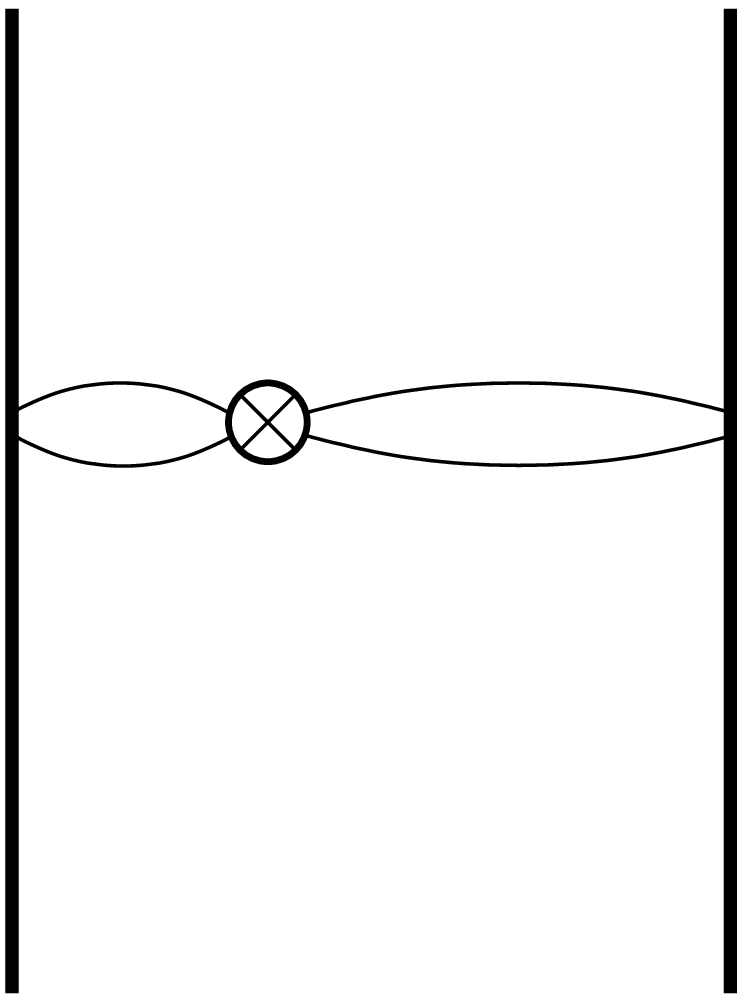}\label{I22}}

\caption{Diagrams representing the different contributions to the
  v.e.v. Solid lines correspond to pairs of particles with opposite
  momenta. Dashed lines correspond to the propagation of zero-momentum
  particles.  The diagrams depicting the terms $D_{ij}$ with $j>i$ can
  be obtained by switching the roles of the two boundaries.}
\label{fig:diag1}
\end{figure}

\begin{figure}[p]
  \centering

\subfigure[$J_{22}$]{\includegraphics[scale=0.4]{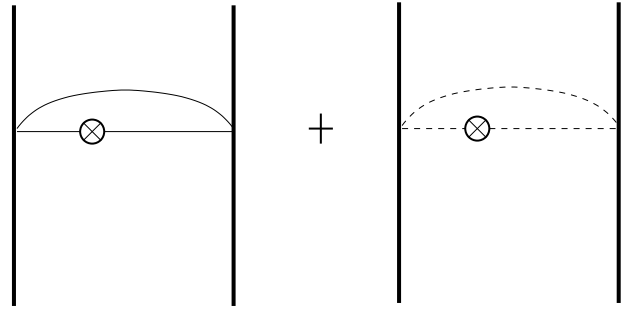}\label{J22}}

\subfigure[$D_{12}$]{\includegraphics[scale=0.4]{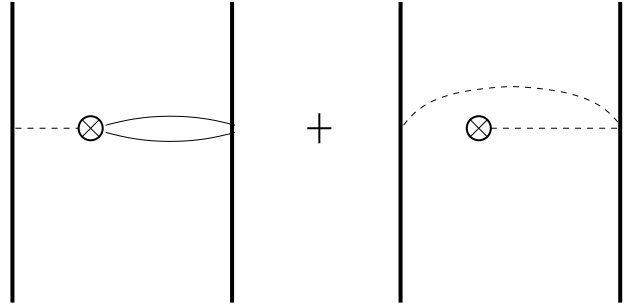}\label{D12}}

\subfigure[$D_{21}$]{\includegraphics[scale=0.4]{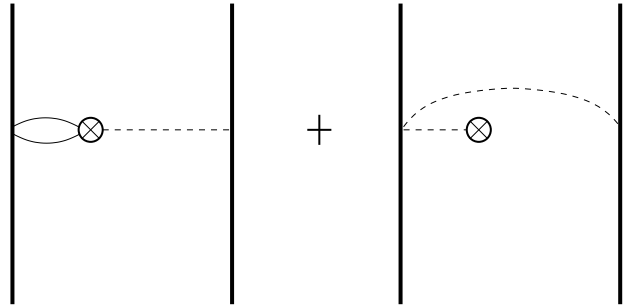}\label{D21}}

\caption{Diagrams representing contributions with singular pieces.
  Solid lines correspond to pairs of particles with opposite
  momentum. Dashed lines correspond to the propagation of
  zero-momentum particles. }
\label{fig:diag2}
\end{figure}

In this appendix we collected the results of section 3. The vacuum
expectation value is given by
\begin{equation}
\begin{split}
  \vev{O(x)}_{R}^{\alpha\beta}=\sum_{i,j}D_{ij}\;.
\end{split}
\end{equation}
We calculated the terms with $i+j\le 4$. The relative
magnitudes of the individual contributions depend on $x$. However, a
consistent ordering can be achieved by putting $x=R/2$. This way
$D_{ij}$ behaves in the large $R$ limit as
$\ordo(e^{-(i+j)mR/2})$. 
A pictorial representation of the individual contributions can be
found in Figs. \ref{fig:diag1} and \ref{fig:diag2}.

\bigskip
{\bf Terms of order $e^{-mR/2}$}

\begin{eqnarray*}
D_{10}&=&\frac{g_\alpha}{2} F_1^O e^{-mx}\\
   D_{01}&=&\frac{g_\beta}{2}  F_1^O e^{-m(R-x)}
\end{eqnarray*}

\bigskip
{\bf Terms of order $e^{-mR}$}

\begin{eqnarray*}
    D_{20}&=& \frac{1}{2}\int\frac{d\theta}{2\pi} K_\alpha(\theta) F_2^O(-\theta,\theta)
  e^{-2m\cosh\theta\ x }\\
 D_{02}&=&  \frac{1}{2}\int\frac{d\theta}{2\pi} K_\beta(\theta) F_2^O(-\theta,\theta)
  e^{-2m\cosh\theta\ (R-x) }\\
  D_{11}&=&\frac{g_\alpha
  g_\beta}{4} F_2^O(i\pi,0)e^{-mR} 
\end{eqnarray*}

\bigskip
{\bf Terms of order $e^{-3mR/2}$}

\begin{eqnarray*}
   D_{30}&=& \frac{1}{2} \int\frac{d\theta}{2\pi} K_\alpha(\theta)  \frac{g_\alpha}{2} F_3^O(-\theta,\theta,0)
  e^{-m(2\cosh\theta+1)\ x }\\
  D_{03}&=& \frac{1}{2} \int\frac{d\theta}{2\pi} K_\beta(\theta)
   \frac{g_\beta}{2} F_3^O(-\theta,\theta,0)
  e^{-m(2\cosh\theta+1)\ (R-x) }\\
  D_{21}&=& \frac{g_\beta}{4}\int\frac{d\theta}{2\pi} 
 \left(F_3^O(-\theta+i\pi,\theta+i\pi,0)K_\alpha(\theta)e^{-2m\cosh\theta  x-m(R-x)}- 
\frac{2(g_\alpha)^2 F_1^O\cosh\theta}{\sinh^2\theta}e^{-m(R+x)} \right)+\\
&& \hspace{6cm}+ e^{-m(x+R)}g_\beta(g_\alpha)^2F_1^O \frac{\varphi(0)}{4}\\
  D_{12}&=& \frac{g_\alpha}{4}\int\frac{d\theta}{2\pi} 
\left(F_3^O(i\pi,-\theta,\theta)K_\beta(\theta)e^{-2m\cosh\theta  (R-x)-mx} - 
\frac{2(g_\beta)^2 F_1^O\cosh\theta}{\sinh^2\theta}e^{-m(2R-x)}  \right)\\
 &&  \hspace{6cm}+
e^{-m(2R-x)}g_\alpha(g_\beta)^2F_1^O \frac{\varphi(0)}{4}
\end{eqnarray*}

\newpage

{\bf Terms of order $e^{-2mR}$}

\begin{eqnarray*}
   D_{40}&=&  \frac{1}{8}\int\frac{d\theta_1}{2\pi}\frac{d\theta_2}{2\pi} 
K_\alpha(\theta_1)K_\alpha(\theta_2) F_4^O(-\theta_1,\theta_1,-\theta_2,\theta_2)
  e^{-2m(\cosh\theta_1+\cosh\theta_2)\ x }\\
   D_{04}&=& \frac{1}{8} \int\frac{d\theta_1}{2\pi}\frac{d\theta_2}{2\pi} 
K_\beta(\theta_1)K_\beta(\theta_2) F_4^O(-\theta_1,\theta_1,-\theta_2,\theta_2)
  e^{-2m(\cosh\theta_1+\cosh\theta_2)\ (R-x)} \\
D_{31}&=&\frac{g_\alpha g_\beta}{8} e^{-mR}  \int \frac{d\theta}{2\pi}  
   K_\alpha(\theta)F_4^O(-\theta+i\pi,\theta+i\pi,i\pi,0)e^{-2m\cosh\theta\ x}\\
D_{13}&=&\frac{g_\alpha g_\beta}{8} e^{-mR}  \int \frac{d\theta}{2\pi}  
   K_\beta(\theta)F_4^O(-\theta+i\pi,\theta+i\pi,i\pi,0)e^{-2m\cosh\theta\ (R-x)}\\
D_{22}&=&I_{22}+J_{22}\text{ , where}\\
I_{22}&=&\frac{1}{4}\int\frac{d\theta_1}{2\pi}\int\frac{d\theta_2}{2\pi}
  K_\alpha(\theta_1)K_\beta(\theta_2)
F_4^O(-\theta_1+i\pi,\theta_1+i\pi,-\theta_2,\theta_2)
e^{-2m\cosh\theta_1 x-2m\cosh\theta_2 (R-x)}\\
&&\text{ and}\\
  J_{22}&=& F_{2}^O(i\pi,0) 
\left\{\int\frac{d\theta}{2\pi}
\Big(K_\alpha(-\theta)K_\beta(\theta)e^{-2m\cosh\theta R}
-\frac{(g_\alpha g_\beta)^2\cosh\theta}{4\sinh^2\theta} e^{-2m R}
\Big)+ \frac{(g_\alpha g_\beta)^2}{8} e^{-2mR}\varphi(0)\right\} 
\end{eqnarray*}

\addcontentsline{toc}{section}{References}
\bibliography{vev_with_boundary}
\bibliographystyle{utphys}

\end{document}